\documentclass[runningheads]{llncs}
\usepackage{tikz}
\usetikzlibrary{calc}
\usepackage{braket}
\newcommand{\cV}{{\cal V}}
\usepackage{amsfonts}
\usepackage{tabularx}

\usepackage{balance}
\usepackage{amsmath,amssymb}
\usepackage{graphicx}
\usepackage{mathrsfs}
\usepackage{tikz}
\usepackage{booktabs}
\usepackage{floatflt}
\usepackage{enumerate}
\usepackage{psfrag}
\usepackage{array}
\usepackage{multirow,hhline}
\usepackage{exscale}
\usepackage{color}
\usepackage{colortbl}
\usepackage{bbm}
\usepackage[ruled,vlined,titlenumbered]{algorithm2e}
\usepackage{braket}
\usepackage{subcaption}

\usetikzlibrary{arrows,decorations.pathmorphing,backgrounds,positioning,fit,petri}
\newcommand{\F}{\mathbb{F}}

\newcommand{\cs}{{L}}
\newcommand{\jc}[1][C]{\ensuremath{\mathfrak{#1 }}}

\usepackage{theorem}
\usepackage{cite}
\usepackage{graphicx}
\newcommand{\ES}[1]{\textcolor{black}{{#1}}}

\newcommand{\cE}{\mathcal{E}}

\newcommand\blfootnote[1]{%
  \begingroup
  \renewcommand\thefootnote{}\footnote{#1}%
  \addtocounter{footnote}{-1}%
  \endgroup
}

\begin{document}
\title{Minimal Trellises for non-Degenerate and  Degenerate Decoding of Quantum Stabilizer Codes}
\titlerunning{Minimal Trellises for Decoding Quantum Stabilizer Codes}
%
\author{Evagoras Stylianou\inst{1} \and
Vladimir Sidorenko\inst{2} \and 
Christian Deppe\inst{3,4} \and Holger Boche\inst{1,4,5,6} }
\authorrunning{E. Stylianou et al.}
%
\institute{ Chair of Theoretical Information Technology,
                    Technical University of Munich 
                    Munich, Germany\\
                    \email{\{evagoras.stylianou, boche\}@tum.de}\and
Institute for Communications Engineering, Technical University of Munich,         Munich, Germany\\
                    \email{vladimir.sidorenko@tum.de} \and
Information Theory and Communication Systems Group, Technical University of Braunschweig, Braunschweig, Germany\\
\email{christian.deppe@tu-braunschweig.de} \and
BMBF Research Hub 6G-life, Germany \and
Munich Center for Quantum Science and Technology (MCQST) \and
Munich Quantum Valley (MQV)
}
\maketitle              

\begin{center}
    \textbf{In Memory of Ning Cai}
\end{center}

\begin{abstract}
This paper presents a comprehensive guide to designing minimal trellises for both non-degenerate and degenerate decoding of quantum stabilizer codes. For non-degenerate decoding, various strategies are explored, leveraging insights from classical rectangular codes to minimize the complexity associated with the non-degenerate maximum likelihood error estimation using the Viterbi algorithm. Additionally, novel techniques for constructing minimal multi-goal trellises for degenerate decoding are introduced, including a merging algorithm, a Shannon-product approach, and the BCJR-Wolf method. The study establishes essential properties of multi-goal trellises and provides bounds on the decoding complexity using the sum-product Viterbi decoding algorithm. These advancements decrease the decoding complexity by a factor ${\cal} O(n)$, where  $n$ is the code length. Finally, the paper applies these results to CSS codes and demonstrates a reduction in complexity by independently applying degenerate decoding to $X$ and $Z$ errors. 
\end{abstract}
%

\blfootnote{This paper is an extension of \cite{evagorasGlobecom}, which will appear in the 2024 IEEE Global Communications Conference (GLOBECOM 2024).}
\section{Introduction}


Quantum computers faced a {major} challenge due to qubit decoherence caused by environmental interactions. Overcoming this obstacle proved difficult as quantum error correction encountered two main hurdles absent in classical error correction: the \emph{no-cloning theorem} \cite{wootters1982single}, which prohibited the duplication of an unknown quantum state, and the collapse of the superposition and destruction of the underlying quantum information upon measuring a quantum state to identify errors.

In his groundbreaking work \cite{shor1995scheme}, Shor illustrated that encoding a single qubit state into a 9-qubit codeword can protect it from general errors. Later, Calderbank, Steane, and Shor \cite{calderbank1996good, steane1996multiple,calderbank1998quantum} devised the CSS and non-CSS constructions, providing a framework for adapting classical error-correcting codes (CECC) to the quantum realm. Gottesman then introduced the stabilizer formalism \cite{gottesman1997stabilizer}, which emerged as the predominant design methodology for quantum error-correcting codes (QECC), facilitating the adaptation of well-known classical code families to the quantum domain \cite{babar2018duality}. One of the principal advantages of this approach is the ability to perform syndrome-based decoding and \ES{this leverages} classical decoding techniques. While successful classical decoding techniques can be employed, decoding stabilizer codes remains challenging due to error degeneracy, where different errors yield the same impact on the code \cite{iyer2015hardness}.

\ES{One of the main Maximum Likelihood (ML) decoding techniques for
classical codes over memoryless channels is the trellis-based Viterbi algorithm \cite{viterbi1967error}.} Although originally used for decoding convolutional codes, trellises were later extended for decoding \ES{block codes} \cite{kschischang1995trellis,mceliece1996bcjr}. \ES{Since the complexity of Viterbi’s algorithm depends} on the number of edges and vertices in the trellis \cite{mceliece1994viterbi}, the concept of a minimal trellis \cite{kschischang1995trellis} is important to control its complexity.
The extension of trellis-based Viterbi decoding to quantum qubit stabilizer codes was pioneered by Ollivier and Tillich \cite{OT2006}. The decoding process is as follows: when a syndrome $\sigma$ is measured, \ES{an error operator $e$ having syndrome $\sigma$ is selected}. Then, the trellis is constructed for the coset $eN$, where all elements in the coset share the syndrome $\sigma$. Here, $N$ represents the normalizer of the stabilizer group ${S}$. Then, by using the Viterbi algorithm for the memoryless depolarizing channel, the (non-degenerate) ML error operator $\hat{e}$ with syndrome $\sigma$ can be found. A modified version of the proposed trellis decoding technique was employed in the decoding of turbo codes in \cite{pelchat2013degenerate,poulin2009quantum}. Subsequently, Xiao and Chen presented a method for transforming stabilizers into the Trellis-Oriented Form (TOF) \cite{XCh2013} while Sabo et al. \cite{sabo2021trellis} further expanded the application of trellis decoding to qudit stabilizer codes (see \cite{ashikhmin2001nonbinary}) and also proposed an alternative algorithm for converting stabilizers to the TOF. Moreover, Sabo et al. \cite{sabo2021trellis}  demonstrated the versatility of trellis decoding for CSS codes by independently decoding the $X$ and $Z$ stabilizers using separate trellises. In another work, Sabo et al. \cite{sabo2022trellis} showcased the utility of trellis decoding in computing weight enumerators and determining the minimum distance of a stabilizer code.



\ES{It is crucial to highlight that the trellis-based Viterbi algorithm is not optimal to decode quantum stabilizer codes in memoryless channels. The
main difference between quantum stabilizer codes and classical codes is degeneracy,} meaning that multiple errors produce the same effect on a codeword, rendering them indistinguishable from each other. These errors lie in the cosets of the stabilizer group $S$ in the normalizer $N$. Consequently, since multiple errors produce the same output, an \emph{optimal} decoder called the \emph{degenerate} ML (DML) decoder, should select the coset of the stabilizer group $S$ \ES{in the normalizer N with the highest probability of occurring. In contrast to the DML, any} \emph{non-degenerate} ML (NDML) decoder, such as the trellis-based Viterbi decoder, opts for the most probable error $e\in N$.{ It is important to mention that the authors in \cite[Section 3.6]{sabo2022trellis} utilized trellises for DML decoding, but the proposed method \emph{did not employ the minimal trellis} for DML decoding. In other words, the \ES{used} trellis did not have the minimal number of vertices and edges. Instead, the authors utilized a trellis for each stabilizer coset in the normalizer.}

The NDML decoder addresses the same problem as the ML decoder for classical codes, which is well-established as NP-complete problem. However, the DML decoder is considerably more complex as it belongs to the $\# \text{P}$-complete class of problems \cite{iyer2015hardness}. Therefore, there exists an inherent trade-off between performance gain and time complexity when choosing the DML decoder over the NDML. It is worth noting that NDML decoding is also an intractable problem, {however, efficient algorithms from classical error correction can be adapted for some code classes.}
 
Efficient implementations of the DML decoder were established for specific \ES{quantum} code families, like quantum convolutional codes \cite{pelchat2013degenerate}, utilizing trellis decoding, concatenated codes \cite{poulin2006optimal}, and Tensor networks \cite{bravyi2014efficient,ferris2014tensor}, which are applicable to 2D quantum codes \cite{chubb2021general} and for the Bacon-Shor codes \cite{napp2012optimal}. However, for a general quantum stabilizer codes, efficient decoding techniques for the DML problem are lacking. In the case of NDML decoding, a prominent general decoding technique is trellis-based decoding \cite{OT2006}, as any stabilizer code admits a trellis representation, although the reduction in complexity depends highly on the code's structure. Other successful NDML decoding techniques have been tailored for specific families of codes, such as the minimum-weight perfect matching 
(MWPM) decoder \cite{fowler2013minimum}, the union-find decoder \cite{delfosse2021almost} for surface codes (for an in-depth exploration of decoding techniques of surface codes see \cite{demarti2023decoding}), and the (enhanced) belief propagation decoder for quantum LDPC codes \cite{babar2015fifteen}.


In this semi-tutorial paper, we focus on trellis-based decoding for general stabilizer code, i.e., we do not assume any structure or specific codes. 
Our investigation unfolds in two parts: firstly, we aim to offer a clear and comprehensive explanation of the trellis-based \ES{NDML decoding for quantum stabilizer codes based on trellises with a single goal node.}  We demonstrate that the properties described in previous literature can be derived \ES{directly from the theory of rectangular codes}. Additionally, we introduce three approaches for constructing stabilizer code trellises. The first approach utilizes the generators of the stabilizer group ${S}$. Contrary to previous claims, we show that the resulting trellis is minimal for any set of generators, regardless of whether they are in the TOF or not. The second approach employs the generators of the normalizer $N$ in the TOF, and finally, we present a method for combining twin nodes in the trivial trellis of a code.

In the second part of our investigation, we propose and analyze the properties of a \emph{minimal} trellis for DML decoding. The proposed trellis is a minimal multi-goal trellis that contains all the cosets of the stabilizer group $S$ within the normalizer $N$. Each coset has its own goal (output) vertex in the trellis. This setup enables the \emph{simultaneous} computation of the probability of each coset using the sum-product \ES{Viterbi algorithm \cite{mceliece1996bcjr}. This approach is able to reduce the complexity} of the DML decoding. We introduce various techniques for constructing the minimal trellis, including a merging algorithm, a BCJR-Wolf method, and an approach based on the Shannon (trellis) product of multi-goal atomic trellises.  Moreover, \ES{we propose a trellis construction tailored specifically to CSS codes} for degenerate decoding using the $X$ and $Z$ generators independently. Finally, we establish complexity bounds for the proposed DML decoder with the minimal multi-goal trellis.


\section{Preliminaries}

In this section, we review quantum stabilizer codes, introduce the concept of code trellises, and elaborate on the decoding of stabilizer codes in memoryless channels.

\subsection{Stabilizer codes}
{Here}, we offer a brief introduction to the stabilizer formalism, for a more detailed exploration see \cite{gottesman1997stabilizer, nielsen2002quantum}.
The single qubit Pauli operators are defined as follows:
\begin{align*} \mathcal{I} = \begin{pmatrix}
1 & 0 \\
0 & 1
\end{pmatrix}, \quad 
\mathcal{X} = \begin{pmatrix}
0 & 1 \\
1 & 0
\end{pmatrix},\quad \mathcal{Z}= \begin{pmatrix}
1 & 0 \\
0 & -1
\end{pmatrix},\quad \mathcal{Y} = \begin{pmatrix}
0 & -i \\
i & 0
\end{pmatrix},
\end{align*}
where $\mathcal{X},\mathcal{Y},\mathcal{Z} \in \mathbb{C}^{2\times 2}$ and $i = \sqrt{-1}$. These operators along with the phases $\{\pm 1, \pm i\}$, generate the single qubit \emph{Pauli group} $\mathcal{P}_1$ under {matrix} multiplication.  This concept is extended to define the \emph{$n$-qubit Pauli group} $\mathcal{P}_n$ denoted by a calligraphic letter, which consists of the $n$-fold tensor product of the single qubit Pauli operators, i.e.,
\begin{align*}
\mathcal{P}_n = \big \{c \cdot B_1 \otimes \cdots \otimes B_n \big|\;B_i \in \{\mathcal{I},\mathcal{X},\mathcal{Y},\mathcal{Z}\},\; c\in \{\pm 1,\pm i\}  \big \}.
\end{align*}
One notable characteristic of $\mathcal{P}_n$ is that its elements either commute or anti-commute. \ES{This is implied by the
property of $P_1$, specifically} $\mathcal{X}^2 = \mathcal{Z}^2 = \mathcal{Y}^2 = \mathcal{I}$ and $\{\mathcal{X},\mathcal{Z}\} = \{\mathcal{X},\mathcal{Y}\} = \{\mathcal{Y},\mathcal{Z}\} = 0$, where $\{\mathcal{A},\mathcal{B}\} = \mathcal{A}\mathcal{B}+ \mathcal{B}\mathcal{A}$.
Since the overall phase $c$ of an operator in $\mathcal{P}_n$ does not have a measurable effect, it is preferable to work with operators in the \emph{effective Pauli group}, defined by (a regular Roman letter) $P_n = \mathcal{P}_n / \{\pm \mathcal{I}^{\otimes n},\pm i \mathcal{I}^{\otimes n}\}$. This means that elements of the effective Pauli group are defined according to equivalence classes, e.g., for $P_1$ (Abelian) we have, $X = [\mathcal{X}] = \{\pm \mathcal{X}, \pm i \mathcal{X}\}$, similarly $Y = [\mathcal{Y}], Z = [\mathcal{Z}]$ and $I = [\mathcal{I}]$ where, 
\begin{align}
XY=Z,\quad XZ=Y, \quad YZ=X,\quad {X}^2 = {Z}^2 = {Y}^2 = {I}.
\label{paulimult}
\end{align}
Since the effective Pauli group $P_n$ is Abelian we can view {the operators} $A_1\otimes A_2\otimes\dots\otimes A_n \in P_n $ as vectors $(a_1,a_2,\dots,a_n)\in P_1^n$ over the single qubit Pauli group $P_1$ with component-wise multiplication {in $P_1$} defined in \eqref{paulimult}. This induces the isomorphism~$P_n\cong P_1^n$. When referring to an element $E\in P_n$ or its vectorized form $e \in P_1^n$ in the effective Pauli group, it corresponds to any element $\mathcal{E}\in \mathcal{P}_n$ with $\mathcal{E}\in[\mathcal{E}]$, that is, with any phase $c\in \{\pm 1,\pm i\}$. Without loss of generality, one can assume that the corresponding element $\mathcal{E}\in \mathcal{P}_n$ of $E\in P_n$ ($e\in P_1^n$) is the element in $[\mathcal{E}]$ with a unity phase, i.e., $c = 1$.

The \emph{stabilizer group} $\mathcal{S}$ is a commutative subgroup of $\mathcal{P}_n$ which does not contain the minus identity operator, i.e., $-\mathcal{I}^{\otimes n}$. {The stabilizer group} can be generated by $n-k$ independent generators denoted by $\mathcal{S}_i$ for $i=1,\dots,n-k$. {The effective stabilizer group is generated by the stabilizer generators $S_i \in P_n$ ($s_i\in P_1^n$) for $i=1,\dots n-k$, which is a subgroup in the effective Pauli group.} The $n$-qubit quantum states are defined as the normalized states $\ket{\psi} = \sum_{x} \alpha_{x}\ket{x}$ where $\sum_x |\alpha_x|^2=1$, $\alpha_x\in \mathbb{C}$ and $\{x\}_{x\in \{0,1\}^n}$ is an orthonormal basis of the Hilbert space $\mathcal{H}=\mathbb{C}^{2^n}$.
For an \emph{$[[n,k]]$ stabilizer code} $\mathcal{Q}$  the codespace is defined as the largest common $+1$ eigenspace of the stabilizer operators,
\begin{align*}
    \mathcal{Q} = \big \{\ket{\psi} \in \mathbb{C}^{2^n}\big | \;\mathcal{S}_i\ket{\psi} = \ket{\psi}, \;\forall i=1,\dots,n-k  \big \}.
\end{align*}
It is important to note here that all states $\ket{\psi}\in \mathcal{Q}$ remain unchanged under multiplication \ES{with any element in the stabilizer group $\mathcal{S}$.} 

\subsection{Code trellises}
\ES{As the goal of the upcoming section is to demonstrate the considered code can be represented as a trellis, we now define and explore some properties of trellises representing block codes.}
A classical block code of length $n$ is a set $C\subseteq A^n$ of words $c=(c_1,c_2,\dots,c_n)$ over the finite alphabet \ES{$A$}. In this section, we do not require that the code or the alphabet have any algebraic structure. All the results in this section are valid for poly-alphabetic codes where $c_i \in A_i$ and the alphabets $A_i$ can be different for $i=1,\dots,n$.
\begin{definition}[Trellis]  \label{def:trellis}
	A trellis of depth $n$  is an edge-labeled directed graph $T=(\mathcal{V},\mathcal{E},A)$ where the vertex set $\mathcal{V}$ is partitioned as $\mathcal{V} = \mathcal{V}_0 \cup \mathcal{V}_1 \cup \dots \cup \mathcal{V}_n$ \ES{such that any edge $\epsilon \in \mathcal{E}$ has the form $\epsilon=(v,v',a)$ with $v\in \mathcal{V}_{i-1},\:v'\in \mathcal{V}_i$, and label $a\in A$ ($A_i$ for poly-alphabetic case). We denote the label of an edge $\epsilon$ by $\lambda(\epsilon)=a$.}
\end{definition} 

\begin{figure}[t]
    \centering




\tikzstyle{state}=[shape=circle,draw=blue!50,fill=blue!20, inner sep=1pt,minimum size=1pt]
\tikzstyle{observation}=[shape=rectangle,draw=orange!50,fill=orange!20]
\tikzstyle{lightedge}=[<-,dotted]
\tikzstyle{mainstate}=[state,thick]
\tikzstyle{mainedge}=[<-,thick]
\tikzstyle{identity}=[<-,solid]
\tikzstyle{Xop}=[<-,dashed, color = red]
\tikzstyle{Zop}=[<-,densely dotted, color = blue]
\tikzstyle{Yop}=[<-,dashdotted, color = green]

\begin{tikzpicture}[]
\node[state] (s1_1) at (0,5) {\tiny $00$};
\node (I) at (0,3) {};
\node[state] (s1_2) at (2,5) {\tiny$00$}
    edge[identity] (s1_1);
\node[state] (s2_2) at (2,4.5) {\tiny$01$}
    edge[Zop] (s1_1);
\node[state] (s3_2) at (2,4) {\tiny$10$}
    edge[Xop] (s1_1);
\node[state] (s4_2) at (2,3.5) {\tiny$11$}
    edge[Yop] (s1_1);
\node (I_edge1) at (1.7,3) {$I$} 
edge[identity] (I);
\node (I_edge2) at (2,3) {};
\node[state] (s1_3) at (4,5) {\tiny$00$}
    edge[identity]  (s1_2)
    edge[Zop] (s2_2)
    edge[Xop] (s3_2)
    edge[Yop] (s4_2);
\node[state] (s2_3) at (4,4.5) {\tiny$01$}
    edge[Zop] (s1_2)
    edge[identity] (s2_2)
    edge[Yop] (s3_2)
    edge[Xop] (s4_2);
\node[state] (s3_3) at (4,4) {\tiny$10$}
    edge[Xop] (s1_2)
    edge[Yop] (s2_2)
    edge[identity] (s3_2)
    edge[Zop] (s4_2);
\node[state] (s4_3) at (4,3.5) {\tiny$11$}
    edge[Yop] (s1_2)
    edge[Xop] (s2_2)
    edge[Zop] (s3_2)
    edge[identity] (s4_2);
\node (X_edge1) at (3.7,3) {$X$} 
edge[Zop] (I_edge2);
\node (X_edge2) at (4,3) {} ;
\node[state] (s1_4) at (6,5) {\tiny$00$}
    edge[identity]  (s1_3)
    edge[Zop] (s2_3)
    edge[Xop] (s3_3)
    edge[Yop] (s4_3);
\node[state] (s2_4) at (6,4.5) {\tiny$01$}
    edge[Zop] (s1_3)
    edge[identity] (s2_3)
    edge[Yop] (s3_3)
    edge[Xop] (s4_3);
\node[state] (s3_4) at (6,4) {\tiny $10$}
    edge[Xop] (s1_3)
    edge[Yop] (s2_3)
    edge[identity] (s3_3)
    edge[Zop] (s4_3);
\node[state] (s4_4) at (6,3.5) {\tiny$11$}
    edge[Yop] (s1_3)
    edge[Xop] (s2_3)
    edge[Zop] (s3_3)
    edge[identity] (s4_3);
\node (Z_edge1) at (5.7,3) {$Z$} 
edge[Xop] (X_edge2);
\node (Z_edge2) at (6,3) {} ;
\node[state] (s1_5) at (8,5) {\tiny $00$}
    edge[identity]  (s1_4)
    edge[Zop] (s2_4)
    edge[Xop] (s3_4)
    edge[Yop] (s4_4);
\node (Y_edge1) at (7.7,3) {$Y$} 
edge[Yop] (Z_edge2);
\end{tikzpicture}

    \caption{\ES{The minimal trellis for the $[[4,2,2]]$ code \cite{vaidman1996error} (see Example~1).}} 
    \label{fig:fourqubit}
\end{figure}
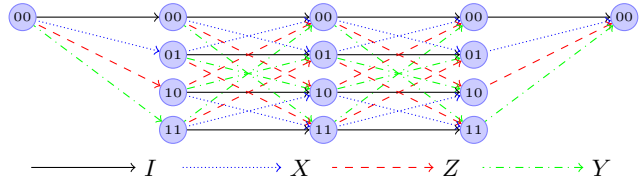
The partition of vertices induces the partition of edges in $\mathcal{E} = \mathcal{E}_1 \cup \mathcal{E}_2 \cup \dots \cup \mathcal{E}_n$ where $\mathcal{E}_i$, called the \emph{trellis section}, consists of all edges in $T$ that end in the vertices of $\mathcal{V}_i$. We assume, unless stated otherwise, that the subsets $\mathcal{V}_0,\mathcal{V}_n$ consist of a single vertex, called the \emph{root}  ($v_{r}$) and the \emph{goal} ($v_{g}$), respectively. A \emph{path} from  $v_{r}$ to $v_{g}$ is the sequence $p= \epsilon_1,\epsilon_2,\dots,\epsilon_n$, where $\epsilon_i\in \mathcal{E}_i$, of connected edges. \ES{If every vertex in $\mathcal{V}$} lies on a path $p$, the trellis is called \emph{reduced}. A trellis is said to be \emph{one-to-one} if all paths in $T$ from $v_{r}$ to $v_{g}$ are labeled distinctly.

An example of a trellis of depth $4$ with the alphabet $A=\{I,X,Y,Z\}$ is shown in Figure 1. 

\ES{The labels of the edges in a path $p$ from  $v_{r}$ to $v_{g}$ in the trellis $T$ of length $n$ defines a word $w(p)= (\lambda(\epsilon_1),\lambda(\epsilon_2),\dots,\lambda(\epsilon_n))$ over the alphabet $A$.}  
\begin{definition}[Code trellis]
	The trellis $T$ of depth $n$ presents a block code $C$ of length $n$ if the set of words $w(p)$  is precisely the set of codewords of $C$. We also say that $T$ is a code trellis of $C$. 
\end{definition}

\begin{definition}[Minimal trellis]
	A trellis $T=(\mathcal{V},\mathcal{E},A)$ presenting the code $C$ is called minimal if it has the minimal number of vertices $|\mathcal{V}|$ among all other \ES{trellises presenting the code.}
\end{definition}

\begin{definition}[{Twin vertices} and biproper trellis]
{Two vertices
	$v',v''\in \cV_t$ are {right (left) twins} if they are connected with a vertex $v\in
	V_{t-1}$ ($v\in V_{t+1}$) by edges with the same label. A trellis without twin vertices is said to be {biproper}. }
 \end{definition}

Given a depth $t\in [1,n-1]$, we split every codeword $c$ into the concatenation of the past $p$ and the future $f$ as  $c=(p,f)$ where $p=(c_1,\dots,c_t)$ and $f=(c_{t+1},\dots,c_n)$. 
\begin{definition}[Rectangular code]
	A classical code $C$ of length $n > 1$ is said to be rectangular if, for all $t \in [1,n-1]$, $\{(a,c), (a,d), (b,c)\} \subset C$ implies that $(b,d)~\in~C$.
\end{definition}

Twin vertices in a trellis presentation of a \emph{rectangular}  code can be merged \cite{kschischang1995trellis}. {Next, consider the following well-known results for trellises in classical coding theory.}
\begin{theorem}[\hspace{-0.005cm}\cite{Kschischang1996},\cite{Sid1997}]\label{Th2Ksch}
For a block code $C$, the following are equivalent.
\begin{enumerate}
	\item  $C$ has a reduced biproper trellis presentation;
	\item  $C$ is rectangular;
	\item  $C$ has a unique minimal trellis presentation.
\end{enumerate}
\end{theorem} 	
\begin{theorem}[\hspace{-0.005cm}\cite{Kschischang1996},\cite{Sid1997}]\label{min}
Among all trellis presentations for a rectangular code, the reduced biproper trellis presentation \ES{minimizes the following quantities:} 
\begin{enumerate}
	\item $|\mathcal{V}|$, total number of vertices; 
	\item $|\mathcal{E}|$, total number of edges;
	\item $|\mathcal{E}|-|\mathcal{V}|$, the cyclic rank of $T$. 
	 \end{enumerate}
 \end{theorem}

\ES{Theorem \ref{Th2Ksch} and \ref{min} imply that by showing a code is rectangular, one automatically shows that it has a unique minimal trellis presentation. Furthermore, showing that the trellis is biproper is sufficient to ensure its minimality.}


\begin{figure}[t]
    \centering
    \includegraphics[width=\columnwidth]{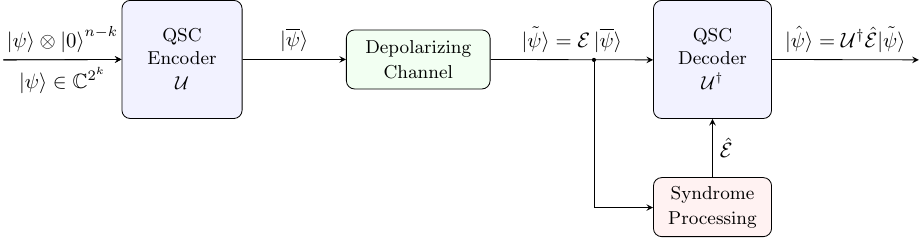}
    \caption{Quantum communications model using QSC.}\vspace{-0.5cm}    \label{fig:com}
\end{figure}

\section{Decoding in Memoryless Channels} \label{sec:depol}
In this section, we illustrate the decoding process for stabilizer codes and distinguish between non-degenerate and degenerate decoding.{ Throughout the rest of the paper, we focus on the memoryless depolarizing channel. For a single qubit state $\rho$, the channel $\mathcal{N}$ acts as follows: 
\begin{align*}
    \mathcal{N}(\rho) = (1-p)\rho + \frac{p}{3}X\rho X + \frac{p}{3}Z\rho Z + \frac{p}{3}Y\rho Y,
\end{align*}
where $p$ is the depolarizing noise. For an $n$-qubit state the memoryless depolarizing channel acts on each qubit independently, that is, the channel is given by $\mathcal{N}_n = \mathcal{N}^{\otimes n}$.} In this channel, the probability of an error $e \in P_1^n$ occurring is expressed as $\text{Pr}(e) = \prod_{i=1}^n \text{Pr}(e_i)$, where $\text{Pr}(\cdot)$ represents a probability function on $P_1$. This implies that errors from $P_1$ can independently occur in any of the $n$ qubits.

In Figure~\ref{fig:com}, we illustrate the quantum communication model under consideration. In this model, a $k$-dimensional quantum state $\ket{\psi} \in \mathbb{C}^{2^k}$ is encoded using a quantum stabilizer code (QSC) encoder into an $n$-dimensional state $\ket{\overline{\psi}}\in \mathbb{C}^{2^n}$. The encoded state is computed by $\ket{\overline{\psi}} = \mathcal{U}(\ket{\psi}\otimes \ket{0}^{n-k})$ where, $\mathcal{U}\in \mathbb{C}^{2^n \times 2^n}$ is the QSC encoder (an isometry) and the zero state $\ket{0}^{n-k}\in \mathbb{C}^{2^{n-k}}$ represents the $n-k$ qubits used for redundancy. This encoded state is then subjected to corruption by the previously mentioned depolarizing channel, resulting in the state $\ket{\tilde{\psi}} = \mathcal{E}\ket{\overline{\psi}}$. Subsequently, we measure the \emph{syndrome} of this corrupted state, \ES{which we will formally define in the next paragraph,} estimate the error $\hat{\mathcal{E}}$, and employ the QSC decoder to recover an estimate $\ket{\hat{\psi}}$ of the original state.

In the context of stabilizer codes, the measurement of the syndrome is determined by the commutative or anti-commutative relationship between the error operator $\mathcal{E}\in \mathcal{P}_n$ and the stabilizers $\mathcal{S}_i$. As we utilize elements of the effective Pauli group, which is Abelian, these properties are lost. To recover them, we utilize the inner product ``$*$'' of elements ${a},{b} \in {P}_1^n$ which is given by ${a} * {b} = \sum_{i=1}^n {a}_i *{b}_i \mod 2$, where,
\begin{align}
{a}_i * {b_i} =    \begin{cases}
        1, & \text{if}\;\;{a}_i \neq {b}_i,\;a_i \neq I, b_i \neq I  \\
    0, & \text{otherwise.}
    \end{cases} .
\end{align}
Clearly, two elements $\mathcal{A},\mathcal{B} \in \mathcal{P}_n$ commute if and only if the corresponding elements $a,b\in P_1^n$ satisfy ${a} * {b} = 0$. If ${c} \in {P}_1$ and ${a} \in {P}_n$, the product ${c}*{a}$ is taken element-wise. Therefore, the syndrome $\sigma({e}) = (\sigma_1,\dots,\sigma_{n-k}) \in \mathbb{F}_2^{n-k}$ of an error vector $e\in P_1^n$ is a binary length $n-k$ vector computed by $$\sigma_i  = {s}_i * {e},\quad \text{for all}\;\;i=1,\dots,n-k.$$

Note that the syndrome {$\sigma({e})$} does not uniquely correspond to an error operator $e \in P_1^n$. \ES{To understand the reason for this, we consider} the effective normalizer group $N$ of the stabilizer group $S$ in $P_1^n$ {defined} as follows:
\begin{align*}
N = \{ {p} \in P_1^n\big | \; {p} * {s} = 0,\;\forall s \in S   \}.
\end{align*}
By definition, all elements in $N$ have a zero syndrome; thus, it follows that $\sigma({e}) = \sigma({be})$ for all $b\in N$. Consequently, given a measured syndrome $\sigma(e)$, the coset $\rho N$, where $\rho$ is any vector in $P_1^n$ with syndrome $\sigma(e)$, contains all vectors that have a syndrome equal to $\sigma(e)$. Therefore, a \emph{non-degenerate} decoder should search for the most probable error in $\rho N$. Formally, the effective Pauli group $P_1^n$ is divided into cosets of the normalizer $N$ and forms the factor group $P_1^n /N$. Since $|P_1^n| = 2^{2n}$ and \ES{$|N| = 2^{n+k}$ there are $2^{n-k}$ cosets where each coset contains all the errors with the same syndrome.} We denote the representatives of these cosets by $\rho\in P_1^n$. 
Given a syndrome $\sigma$, we can choose the coset $\rho N$, \ES{and can then express the NDML decoding as:}
\begin{align*}
\hat{e}_{{\mathrm{NDML}}} = \arg\max_{e \in \rho N } \mathrm{Pr}(e).
\end{align*}
In memoryless channels, the error probability is simplified to $\mathrm{Pr}(e)=\prod_{i=1}^n \mathrm{Pr}(e_i)$. The NDML decoder aligns with classical decoding approaches for error correcting codes, \ES{allowing to adapt classical techniques to the quantum setting. In this
paper, we demonstrate how the trellis-based Viterbi algorithm, originally
introduced in \cite{OT2006} can be used for NDML decoding.}


The NDML decoding for the memoryless depolarizing channel can be summarized as follows:\\
(1) Given the syndrome $\sigma$, choose  a representative $\rho$ of the coset $\rho N$.\\ 
(2) Compute the probability  $\text{Pr}(e)$ of each $e\in \rho N$, using the formula
\begin{equation*}\label{eq:p(b)}
\text{Pr}(e)= \prod_{i=1}^n \text{Pr}(e_i).
\end{equation*}
(3) Output the vector $e$ with the largest probability~$\text{Pr}(e)$. \ES{If there are multiple candidates, choose one at random.}

The algorithm's complexity is primarily governed by Step 2, necessitating $n2^{n+k}$ multiplications. To mitigate the complexity of the NDML decoder, we demonstrate the construction of the trellis for the group code $N$. It is worth noting that the trellises of the cosets $\rho N$ can be computed by relabeling the trellis of $N$.

However, due to the unique features of quantum stabilizer codes compared to classical codes, the above technique does not compute the optimal error in the ML sense.  \ES{In fact, for any quantum} stabilizer code $\mathcal{Q}$ there exists distinct errors $\mathcal{E},\mathcal{E}' \in \mathcal{P}_n$ that have the same effect on the codes, that is,
\begin{align*}
    \mathcal{E}\ket{\psi} = \mathcal{E}'\ket{\psi},\;\; \text{for all}\;\ket{\psi} \in \mathcal{Q}.
\end{align*}
By definition, any stabilizer element $\mathcal{A} \in \mathcal{S}$ has no effect on the code. \ES{This implies that all errors produce the same} output differ only by a stabilizer element, i.e., $\mathcal{E} = \mathcal{A}\mathcal{E}'$ for some  $\mathcal{A}\in \mathcal{S}$. In terms of the vector group $P_1^n$, we have that all vectors $e,e'\in P_1^n$ are such that $e = e's $ for some $s\in S$. Since, both errors have the same syndrome, say $\sigma$ with a representative $\rho$, but differ up to a stabilizer \ES{element, they must be in one of the cosets} of the stabilizer $S$ in the normalizer coset $\rho N$. Since, $|N| = 2^{n+k}$ and $|S|=2^{n-k}$, \ES{there are} $2^{2k}$ such cosets. We denote the group of the representatives of these cosets by $L$ and its generators by $\{\ell_i\}_{i=1}^{{2k}},\:\ell_i \in P_1^n$ and we {refer to its elements} as logical operators.  \ES{They are called logical operators, as they map a state in the code to another state in the code.}


Since all the elements of the cosets of $S$ in $\rho N$ have the same effect on the code, the DML decoder should choose the coset with the largest probability. By the above factorization, any error $e\in P^n_1$ can be decomposed as $e = \rho \ell s $ where $\rho $ is the representative of the coset $\rho N$, $\ell \in L$ and $s \in S$. By measuring the syndrome $\sigma$ we can choose $\rho$ and since $s$ has no effect, the DML decoder can be expressed~as: 
\begin{align*}
    \hat{\ell}_{{\mathrm{DML}}} = \arg\max_{\ell \in L} \mathrm{Pr}(\ell| \sigma ),
\end{align*}
where the conditional probability $\mathrm{Pr}(\ell |\sigma)$ can be computed \ES{by summing the probabilities} of all elements in the coset $\ell \rho S$. Therefore, the DML decoding reduces to:
\begin{align*}
\hat{\ell}_{{\mathrm{DML}}} = \arg\max_{\ell \in L} \sum_{s\in S} \mathrm{Pr}(\rho \ell s).
\end{align*}

The DML decoding is then summarized as follows: \\ 
(1)  Given the syndrome $\sigma$, find $\rho$ such that the coset $\rho N$ has the syndrome $\sigma$.\\ 
{(2) Compute the probability  $\text{Pr}(\ell)$ of each coset $\rho \ell S$ {in $\rho N$} for all $\ell \in L$, using the following \emph{sum-product} formula.}
  \begin{equation*}\label{eq:p(b)}
\text{Pr}(\ell)= \sum_{w\in \rho \ell S} \prod_{i=1}^n \text{Pr}(w_i)
= \sum_{w\in \ell S} \prod_{i=1}^n \text{Pr}(\rho_i w_i).
\end{equation*}
(3) Output $\ell$ \ES{with the largest} probability~$\text{Pr}(\ell)$. 


{This decoding relies on the cosets of $S$ in $N$, and not directly on the cosets of $S$ in $\rho N$, as the representatives $\rho$ are only present in the probability $\text{Pr}(\rho_i w_i)$.}

The algorithm's complexity is dominated by Step 2, which requires $n2^{n+k}$  multiplications and $2^{n+k}$ additions.  \ES{We can reduce the cost by using} the minimal multi-goal trellis $T$ {that represents all cosets of $S$ in $N$ in one trellis.}

\section{Minimal trellises for non-Degenerating Decoding}

In this section, we illustrate a method for reducing the complexity of the NDML decoder by utilizing code trellises. Our focus lies in demonstrating that existing methods for constructing trellises for classical codes can be directly applied to NDML decoding of quantum stabilizer codes. Consequently, there is no need to rederive the properties of such trellises, such as demonstrating their minimality.
\ES{Recall that the NDML decoder aims at identifying the most} probable error in $\rho N$ after measuring the syndrome $\sigma$ and determining $\rho$. Going forward, our focus shifts exclusively to the normalizer $N$ instead of the cosets $\rho N$. \ES{We will show that all properties} and methods apply to any cosets through trellis relabeling. 

We can define the normalizer (the block code) $N$ of length $n$ as follows. Given the generators $s_1,\dots,s_{n-k}\in P_1^n$ (in vector form), of the stabilizer group $S$, the code $N$ consists of all vectors $c$ that are orthogonal to all stabilizers in $S$, i.e.,
\begin{equation}\label{norm_code}
N=\{c\in P_1^n | c*s_i = 0, \;\;\forall i\in [1,n-k]\}.
    \end{equation}
The code $N$ is an Abelian multiplicative group under component-wise multiplication of the codewords and $|N|=2^{n+k}$.

The code  $N$ is \emph{rectangular} since it is a group code. Indeed, let us consider codewords $c_1, c_2, c_3 \in N$, and we split them into past and future for a fixed \ES{index} $t$. Specifically, $c_1 = (a, d)$, $c_2 = (a, c)$, and $c_3 = (b, c)$. Then, the product $c_1 c_2^{-1} c_3 = (a,d)(a^{-1},c^{-1})(b,c) = (b,d)$ must be a codeword, since $N$ is a group. Therefore, since the code $N$ is rectangular, according to Theorem~\ref{Th2Ksch}, it has a unique minimal reduced biproper trellis. Additionally, for rectangular codes, any classical method for designing a minimal trellis can be directly applied to $N$.

Assume that we have measured the syndrome $\sigma(e)$ of the error $e\in P_1^n$, corresponding to the representative $\rho$. To find the most probable error vector in the coset $\rho N$ with this syndrome, we assign a weight of $-\log \text{Pr}(\lambda(\epsilon))$ to every edge $\epsilon$ of the trellis representing $\rho N$. We then employ the standard Viterbi algorithm \cite{viterbi1967error} to identify a path in the trellis with the minimum weight.

The Viterbi decoder entails $|\mathcal{V}|$ additions and $|\mathcal{E}|-|\mathcal{V}|+1$ binary selections. Thus, according to Theorem~\ref{min}, utilizing the minimal trellis yields the minimum decoding complexity. To compute the aposteriori probability of each error component separately, one can employ the standard BCJR algorithm \cite{bahl1974optimal}, which exhibits a complexity three times higher than that of the Viterbi algorithm. Therefore, adopting the minimal trellis reduces the complexity of the BCJR algorithm as well.

We will now proceed to illustrate how to construct the minimal trellises.

\subsection{Using the generators of stabilizer group $S$} \label{sec:wolfnorm}

Let $H(N)$ be the $(n-k)\times n$ matrix over the alphabet $P_1$ that contains as rows $n-k$ generators of the stabilizer group $S$ as follows:
\begin{align*}
  H(N) = \begin{pmatrix}
      s_1\\
      s_2\\
      \vdots\\
      s_{n-k}
  \end{pmatrix}  = \begin{pmatrix}
      h_1 & h_2 &\cdots & h_n
  \end{pmatrix}.
\end{align*}
The syndrome of an $n$-vector
$c=(c_1,\dots,c_n)\in P_1^n$ is the binary $(n-k)$-vector $\sigma\in \F_2^{n-k}$ defined by
\begin{equation}\label{eq:syndr}
\sigma(c) = c*H(N)^T= \sum_{i=1}^n  c_i * h_i.  
\end{equation} 
The code $N$ is defined as the set of all vectors from $P_1^n$ that have the zero syndrome, that is,
\begin{equation*}
N=\{c\in P_1^n | c*H(N)^T = 0\}.
\end{equation*}
Hence, we can say that $H(N)$ is a \emph{check matrix} for the group code $N$.
The partial syndrome of $c$ for time $t\in [1,n]$ is defined as follows:
\begin{equation}
    \sigma_t(c) =  \sum_{i=1}^t  c_i * h_i.
\end{equation}
\begin{algorithm}[t]
	Input: check matrix $H(N)$ of the code $N$;\\
	For every $t=0,1,\dots,n-1$ do\\
	\quad For every vertex $v\in \mathcal{V}_t$ do\\
	\quad\quad For every symbol $p\in P_1$ do\\
	\quad\quad\quad draw an edge $\epsilon$ from $v$ to vertex $v'\in \mathcal{V}_{t+1}$\\ 
	\quad\quad\quad with $f(v')= f(v)+p*h_{t+1}$\\
	\quad\quad\quad  and put label $\lambda(v')=p$.\\
	Output: the complete trellis $T$ for the code $N$.
	\caption{The BCJR-Wolf algorithm}\label{alg:BCJR}
\end{algorithm}

The BCJR-Wolf algorithm, given in Algoritm~\ref{alg:BCJR}, constructs a trellis $T$ for the code $N$ as follows: At each depth $t$ of $T$, the vertices $v\in \mathcal{V}_t$ are enumerated by the binary vectors $f(v)\in \mathbb{F}_2^{n-k}$. \ES{Every vector $w\in P_1^n$ is represented by the path in $T$ that goes through the vertex at depth $t$ that has label equal to its partial syndrome $\sigma_t(w)$ for $t=0,\dots, n$. }\ES{Therefore, the code $N$ consists of all the paths} originating from the initial vertex $v_r \in V_0$ and terminating at the vertex $v \in \cV_n$ in the final section labeled by $f(v) = 0$, in other words, all paths representing vectors in $P_1^n$ with zero syndrome.

Moreover, the BCJR-Wolf method presents all cosets of the code $N$ in $P_1^n$, i.e., the cosets of vectors with the same syndrome. Therefore, if a coset has syndrome $\sigma$, then the presentation of that coset consists of all paths that start from the initial vertex and terminate at the vertex $v\in \mathcal{V}_n$ with label $f(v) = \sigma$. Hence, the BCJR-Wolf method constructs the \emph{complete} trellis of $N$, \ES{also called} the \emph{multi-goal} trellis of all cosets of $P_1^n$.

To obtain the minimal trellis only for the code $N$, one can remove all paths that do not end in a goal node $v\in \cV_n$ with $f(v) = 0$. Then, any coset of the normalizer $\rho N$ \ES{(where $\rho$ is chosen based on the syndrome)} can be constructed by multiplying the labels of the edges $\epsilon \in \mathcal{E}_i$ of each section $i=1,\dots,n$ by the corresponding element $\rho_i$.
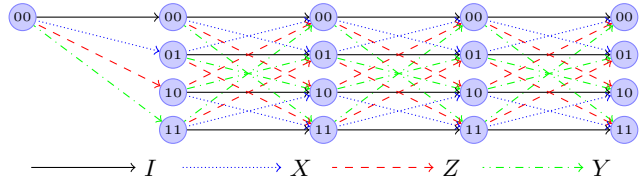
\begin{figure}[t!]
    \centering




\tikzstyle{state}=[shape=circle,draw=blue!50,fill=blue!20, inner sep=1pt,minimum size=1pt]
\tikzstyle{observation}=[shape=rectangle,draw=orange!50,fill=orange!20]
\tikzstyle{lightedge}=[<-,dotted]
\tikzstyle{mainstate}=[state,thick]
\tikzstyle{mainedge}=[<-,thick]
\tikzstyle{identity}=[<-,solid]
\tikzstyle{Xop}=[<-,dashed, color = red]
\tikzstyle{Zop}=[<-,densely dotted, color = blue]
\tikzstyle{Yop}=[<-,dashdotted, color = green]

\begin{tikzpicture}[]
\node[state] (s1_1) at (0,5) {\tiny $00$};
\node (I) at (0,3) {};
\node[state] (s1_2) at (2,5) {\tiny$00$}
    edge[identity] (s1_1);
\node[state] (s2_2) at (2,4.5) {\tiny$01$}
    edge[Zop] (s1_1);
\node[state] (s3_2) at (2,4) {\tiny$10$}
    edge[Xop] (s1_1);
\node[state] (s4_2) at (2,3.5) {\tiny$11$}
    edge[Yop] (s1_1);
\node (I_edge1) at (1.7,3) {$I$} 
edge[identity] (I);
\node (I_edge2) at (2,3) {};
\node[state] (s1_3) at (4,5) {\tiny$00$}
    edge[identity]  (s1_2)
    edge[Zop] (s2_2)
    edge[Xop] (s3_2)
    edge[Yop] (s4_2);
\node[state] (s2_3) at (4,4.5) {\tiny$01$}
    edge[Zop] (s1_2)
    edge[identity] (s2_2)
    edge[Yop] (s3_2)
    edge[Xop] (s4_2);
\node[state] (s3_3) at (4,4) {\tiny$10$}
    edge[Xop] (s1_2)
    edge[Yop] (s2_2)
    edge[identity] (s3_2)
    edge[Zop] (s4_2);
\node[state] (s4_3) at (4,3.5) {\tiny$11$}
    edge[Yop] (s1_2)
    edge[Xop] (s2_2)
    edge[Zop] (s3_2)
    edge[identity] (s4_2);
\node (X_edge1) at (3.7,3) {$X$} 
edge[Zop] (I_edge2);
\node (X_edge2) at (4,3) {} ;
\node[state] (s1_4) at (6,5) {\tiny$00$}
    edge[identity]  (s1_3)
    edge[Zop] (s2_3)
    edge[Xop] (s3_3)
    edge[Yop] (s4_3);
\node[state] (s2_4) at (6,4.5) {\tiny$01$}
    edge[Zop] (s1_3)
    edge[identity] (s2_3)
    edge[Yop] (s3_3)
    edge[Xop] (s4_3);
\node[state] (s3_4) at (6,4) {\tiny $10$}
    edge[Xop] (s1_3)
    edge[Yop] (s2_3)
    edge[identity] (s3_3)
    edge[Zop] (s4_3);
\node[state] (s4_4) at (6,3.5) {\tiny$11$}
    edge[Yop] (s1_3)
    edge[Xop] (s2_3)
    edge[Zop] (s3_3)
    edge[identity] (s4_3);
\node (Z_edge1) at (5.7,3) {$Z$} 
edge[Xop] (X_edge2);
\node (Z_edge2) at (6,3) {} ;
\node[state] (s1_5) at (8,5) {\tiny$00$}
    edge[identity]  (s1_4)
    edge[Zop] (s2_4)
    edge[Xop] (s3_4)
    edge[Yop] (s4_4);
\node[state] (s2_5) at (8,4.5) {\tiny$01$}
    edge[Zop] (s1_4)
    edge[identity] (s2_4)
    edge[Yop] (s3_4)
    edge[Xop] (s4_4);
\node[state] (s3_5) at (8,4) {\tiny $10$}
    edge[Xop] (s1_4)
    edge[Yop] (s2_4)
    edge[identity] (s3_4)
    edge[Zop] (s4_4);
\node[state] (s4_5) at (8,3.5) {\tiny$11$}
    edge[Yop] (s1_4)
    edge[Xop] (s2_4)
    edge[Zop] (s3_4)
    edge[identity] (s4_4);
\node (Y_edge1) at (7.7,3) {$Y$} 
edge[Yop] (Z_edge2);
\end{tikzpicture}

    \caption{Complete (\emph{multi-goal}) minimal trellis of the code $N$ ($P_1^n / N$) from Ex.~\ref{ex:BCJR}.}  
    \label{fig:BCJR}
\end{figure}

\begin{theorem}
	\ES{The BCJR-Wolf Algorithm (Algorithm 1) outputs} the minimal code trellis.
\end{theorem}
\begin{proof}
Algorithm 1 ensures that the obtained complete trellis $T$ is biproper \cite[Th. 7]{Kschischang1996}). Since the code trellis  is a subtrellis  of $T$, it inherits the property of being biproper. Therefore, by Theorem~\ref{Th2Ksch}, the code trellis is minimal.
\end{proof}

\ES{It is noticeable that}, contrary to claims in previous literature \cite{OT2006}, Algorithm~\ref{alg:BCJR} constructs the minimal trellis using \emph{any} check matrix $H(N)$ of the code~$N$.

\begin{example}\label{ex:BCJR} 
The stabilizer group (in vector form) of the 4-qubit code $[[4,2,2]]$ given in \cite{vaidman1996error} is given by $S = \langle XXXX, ZZZZ \rangle$, where $\langle \cdot\rangle$ denotes that the group is generated by the multiplication of the elements (vectors) inside the brackets. Consequently, a parity check matrix of $N$ is given by:
\begin{align*}
H(N)=
\left(
\begin{array}{cccc}
X&X&X&X\\
Z&Z&Z&Z
\end{array}
\right).
\end{align*}
After executing Algorithm~\ref{alg:BCJR}, we obtain the complete trellis $T$ as depicted in Figure~\ref{fig:BCJR}. It's noticeable that the complete trellis has four goal nodes labeled with all the two-bit binary vectors. The paths terminating in the goal node labeled by $(0,0)$ exactly present the code $N$. Furthermore, all paths ending in the goal node labeled by \ES{$(a,b)$ present all the vectors in $P_1^n$ with a syndrome equal to $(a,b)$.} The constructed trellis can be directly utilized in the Viterbi algorithm to decode $N$ (or a particular coset) using different goal vertices. We can obtain the minimal trellis presentation for the code $N$ or for a particular coset by removing all vertices and edges that are not used. Figure \ref{fig:fourqubit} displays the trellis for the code $N$ obtained using the goal vertex numbered by $(0,0)$.
\end{example}

\subsection{Using generators of normalizer set $N$}
\ES{In the previous section, we utilized} the generators of the stabilizer group $S$, \ES{instead we will now employ the
generator} of the normalizer $N$. Let $g_1, \dots, g_{n+k} \in P_1^n$ be the generators of the normalizer code $N$ in their vector form, i.e., $N = \langle g_1, \dots, g_{n+k} \rangle$. Consider the $(n+k) \times n$ matrix $G(N)$ having the generators $g_i,\;i=1,\dots,n+k$ as rows:
\begin{align*}
G(N) = \begin{pmatrix}
g_1\\
g_2\\
\vdots\\
g_{n+k}
\end{pmatrix} \in P_1^{(n+k) \times n}.
\end{align*}

To obtain the minimal trellis $T(N)$, we must transform the generator matrix $G(N)$ into the left-right (LR) form $G'(N)$, also known as the TOF. Note that, unlike the previous construction, not every generator matrix $G(N)$ is suitable for computing the minimal trellis of the code $N$.

Consider a row $g$ of $G$. We define the left index $L(g)$ as the minimum index $t$ such that $g_t \ne I$. Similarly, the right index $R(g)$ is defined as the maximum index $t$ such that $g_t \ne I$. The span of a row $g$ is the interval $[L(g), R(g)]$, and its span length is $R(g) - L(g) + 1$. Finally, the span length of the matrix $G$ is the sum of the span lengths of its rows. 

We then say that a matrix is in the TOF according to the following definition, which is similar to \cite{XCh2013, sabo2021trellis }:
\begin{definition}[TOF or LR-matrix]\label{def:LR}
	A matrix $G$ is in the TOF (is an $LR$-matrix) if for every index $t\in[1,n]$ it has at most two rows $g_{i}$ and $g_j$  that have the same left (or right) index $t$ and $g_{i,t} \ne g_{j,t}$. 
\end{definition}
If only the left (right) indices satisfy the condition in the above definition, we refer to the matrix as having the $L$-property ($R$-property, respectively). We later give an algorithm to transform any matrix $G$ to its TOF.

The following construction of the minimal trellis is based on combining or, more precisely, taking the product of trellises presenting smaller codes. Hence, we need to define how to compute the product of two trellis presentations $T(C')$ and $T(C'')$ presenting the codes $C'$ and $C''$, respectively, of length $n$ over~$P_1^n$.

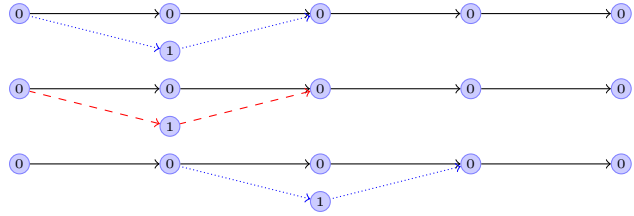
\begin{figure}[t]
    \centering




\tikzstyle{state}=[shape=circle,draw=blue!50,fill=blue!20, inner sep=1pt,minimum size=1pt]
\tikzstyle{observation}=[shape=rectangle,draw=orange!50,fill=orange!20]
\tikzstyle{lightedge}=[<-,dotted]
\tikzstyle{mainstate}=[state,thick]
\tikzstyle{mainedge}=[<-,thick]
\tikzstyle{identity}=[<-,solid]
\tikzstyle{Xop}=[<-,dashed, color = red]
\tikzstyle{Zop}=[<-,densely dotted, color = blue]
\tikzstyle{Yop}=[<-,dashdotted, color = green]

\begin{tikzpicture}[]
\node[state] (s1_1) at (0,5) {\tiny $0$};
\node (I) at (0,3) {};
\node[state] (s1_2) at (2,5) {\tiny$0$}
    edge[identity] (s1_1);
\node[state] (s2_2) at (2,4.5) {\tiny$1$}
    edge[Zop] (s1_1);
\node (I_edge2) at (2,3) {};
\node[state] (s1_3) at (4,5) {\tiny$0$}
    edge[identity]  (s1_2)
    edge[Zop] (s2_2);
\node[state] (s1_4) at (6,5) {\tiny$0$}
    edge[identity]  (s1_3);
\node[state] (s1_5) at (8,5) {\tiny $0$}
    edge[identity]  (s1_4);

\node[state] (s1_1_z) at (0,4) {\tiny $0$};
\node[state] (s1_2_z) at (2,4) {\tiny$0$}
    edge[identity] (s1_1_z);
\node[state] (s2_2_z) at (2,3.5) {\tiny$1$}
    edge[Xop] (s1_1_z);
\node[state] (s1_3_z) at (4,4) {\tiny$0$}
    edge[identity]  (s1_2_z)
    edge[Xop] (s2_2_z);
\node[state] (s1_4_z) at (6,4) {\tiny$0$}
    edge[identity]  (s1_3_z);
\node[state] (s1_5_z) at (8,4) {\tiny $0$}
    edge[identity]  (s1_4_z);

\node[state] (s1_1_3) at (0,3) {\tiny $0$};
\node[state] (s1_2_3) at (2,3) {\tiny$0$}
    edge[identity] (s1_1_3);
\node[state] (s1_3_3) at (4,3) {\tiny$0$}
    edge[identity]  (s1_2_3);

\node[state] (s2_3_3) at (4,2.5) {\tiny$1$}
    edge[Zop] (s1_2_3);    

\node[state] (s1_4_3) at (6,3) {\tiny$0$}
    edge[identity]  (s1_3_3)
    edge[Zop] (s2_3_3);
    
\node[state] (s1_5_3) at (8,3) {\tiny $0$}
    edge[identity]  (s1_4_3);
\end{tikzpicture}
        \caption{Minimal trellis presentation of $g_1 = XXII$, $g_2 = ZZII$ and $g_3 = IXXI$ of the normalizer code $N$.}
        \label{fig:shannon_prod}
\end{figure}

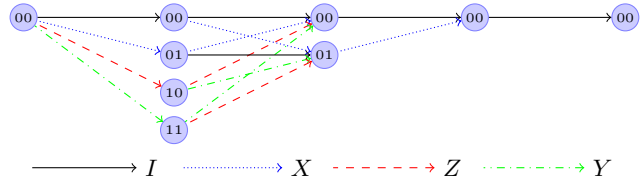
\begin{figure}[t]
        \centering




\tikzstyle{state}=[shape=circle,draw=blue!50,fill=blue!20, inner sep=1pt,minimum size=1pt]
\tikzstyle{observation}=[shape=rectangle,draw=orange!50,fill=orange!20]
\tikzstyle{lightedge}=[<-,dotted]
\tikzstyle{mainstate}=[state,thick]
\tikzstyle{mainedge}=[<-,thick]
\tikzstyle{identity}=[<-,solid]
\tikzstyle{Xop}=[<-,dashed, color = red]
\tikzstyle{Zop}=[<-,densely dotted, color = blue]
\tikzstyle{Yop}=[<-,dashdotted, color = green]

\begin{tikzpicture}[]
\node[state] (s1_1) at (0,5) {\tiny $00$};
\node (I) at (0,3) {};
\node[state] (s1_2) at (2,5) {\tiny$00$}
    edge[identity] (s1_1);
\node[state] (s2_2) at (2,4.5) {\tiny$01$}
    edge[Zop] (s1_1);
\node[state] (s3_2) at (2,4) {\tiny$10$}
    edge[Xop] (s1_1);
\node[state] (s4_2) at (2,3.5) {\tiny$11$}
    edge[Yop] (s1_1);
\node (I_edge1) at (1.7,3) {$I$} 
edge[identity] (I);
\node (I_edge2) at (2,3) {};
\node[state] (s1_3) at (4,5) {\tiny$00$}
    edge[identity]  (s1_2)
    edge[Zop] (s2_2)
    edge[Xop] (s3_2)
    edge[Yop] (s4_2);
\node[state] (s2_3) at (4,4.5) {\tiny$01$}
    edge[Zop] (s1_2)
    edge[identity] (s2_2)
    edge[Yop] (s3_2)
    edge[Xop] (s4_2);
\node (X_edge1) at (3.7,3) {$X$} 
edge[Zop] (I_edge2);
\node (X_edge2) at (4,3) {} ;
\node[state] (s1_4) at (6,5) {\tiny$00$}
    edge[identity]  (s1_3)
    edge[Zop] (s2_3);
\node (Z_edge1) at (5.7,3) {$Z$} 
edge[Xop] (X_edge2);
\node (Z_edge2) at (6,3) {} ;
\node[state] (s1_5) at (8,5) {\tiny $00$}
    edge[identity]  (s1_4);
\node (Y_edge1) at (7.7,3) {$Y$} 
edge[Yop] (Z_edge2);
\end{tikzpicture}

        \caption{Shannon product of the minimal trellises of $g_1 = XXII$, $g_2 = ZZII$ and $g_3 = IXXI$.} \vspace{-0.5cm}
         \label{fig:shannon_prod2}
\end{figure}

\begin{definition}[Shannon product \cite{kschischang1995trellis},\cite{Sidorenko1996}] 
	Given two trellises $T'$ and $T''$, for $t\in [1,n]$ the $t$-th trellis section $\mathcal{E}_t$ of the Shannon product $T=T'\times T''$ is the Cartesian product $\mathcal{E}_t=\mathcal{E}'_t \times \mathcal{E}''_t$ where an element 
	$(v'_1,v'_2,\alpha'),(v''_1,v''_2,\alpha'')$ in this product is an edge 
	$(v_1,v_2,\alpha)=\left( (v_1',v_1''),  (v_2',v_2''), \alpha'\alpha''\right) $ in $T$. 
\end{definition}
For an example of the Shannon product, see  Figure~\ref{fig:shannon_prod} and Figure~\ref{fig:shannon_prod2}. Consider two codes $C'$ and $C''$ of length $n$ over $P_1^n$. The product of the two codes is denoted by $C' \times C''$, and defined as follows: 
\begin{equation}
	C' \times C'' = \{c' c'' | c'\in C', c''\in C'' \}. \label{product}
\end{equation} 
\begin{lemma}[\hspace{-0.05cm}\cite{kschischang1995trellis},\cite{Sidorenko1996}]\label{L1}
	The Shannon product is commutative, and the product of trellises  $T(C')$ and $T(C'')$ is a trellis of the product code $C'\times C''$.
\end{lemma}

Let $T(G(N))$ denote a trellis of the code $N$ with the generator matrix $G$. To construct such a trellis, we first construct the minimal trellis for every "atomic" code $C(g_i)$, which consists of only two words: $g_i$ and the all-identity word $(I, I, \dots, I)$. We refer to the minimal trivial trellis presentation of the atomic codes as the "atomic" trellis.


\begin{example} Consider the $[[4,2,2]]$ code. The following generator matrix of $N$ satisfies the left-right property, i.e., it is in the TOF:
\begin{align*}
G(N)= \begin{pmatrix}
X&X&I&I\\
Z&Z&I&I\\
I&X&X&I\\
I&Z&Z&I\\
I&I&X&X\\
I&I&Z&Z\\ 
\end{pmatrix}.
\end{align*}
The minimal trellises for the first three generators are shown in Figure~\ref{fig:shannon_prod}. The Shannon product of these trellises can be found in Figure~\ref{fig:shannon_prod2}. It is easy to verify that the resulting trellis in Figure~\ref{fig:shannon_prod2} presents the product code $C(g_1)\times C(g_2)\times C(g_3)$.
\end{example}

We are now ready to demonstrate that the Shannon product construction of minimal trellises produces the minimal trellis of a code.

\begin{theorem}
	Given a matrix $G$ in the TOF, the trellis $T(G)$ obtained by the Shannon product of the minimal trellises $T(g_i)$ is minimal.
\end{theorem}
\begin{proof}

By Lemma \ref{L1}, the trellis $T(G(N))$ is a trellis of the code $N$. \ES{We will now show minimality by showing that T (G) is biproper.}

\ES{By contradiction, suppose} that there exist two vertices $v'$ and $v'' \in \mathcal{V}_t$ that are right twins. This scenario can only occur if either 1) there are no rows in $G$, say $g_1$ and $g_2$, with the same left index $t$ and $g_{1,t} = g_{2,t}$, or 2) there are three or more rows in $G$ with the same left index $t$.

In the first case, there would be $2^2=4$ edges outgoing from $v$ labeled with the 2 symbols $g_{1,t}$ and $g_{1,t}g_{1,t}=I$, leading to edges with the same label. In the second case, there would be at least $2^3 = 8$ edges outgoing from $v$ labeled with at most $4$ labels $I,X,Y,Z$, leading again to edges with the same label. However, since $G$ is an $LR$-matrix, both of these cases are impossible.

Similarly, the trellis $T(G)$ has no right twins, establishing that $T(G)$ is a biproper trellis. By Theorem~\ref{min}, the trellis $T(G)$ is minimal.
\end{proof}

\begin{algorithm}[t]
    Input: matrix $G$\\
	Repeat: Search for two or three rows that satisfy the conditions (a) or (b)\\
		\quad If found,\\ 
		\quad\quad Apply the corresponding replacement in (a) or in (b) \ES{and repeat} \\
		\quad else\\ \quad \quad \ES{If no twins are found, output $G$ and stop. }
\caption{Convert a matrix $G$ in the TOF (LR-form)}\label{algLR}		
\end{algorithm}

We introduce the Greedy algorithm (Algorithm \ref{algLR}) to transform a matrix into its TOF. In this algorithm, we examine the following two conditions and take their respective actions:\\
\emph{Condition} (a): Two rows in \( G \), denoted as \( g_1 \) and \( g_2 \), share the same left (right) index, and \( g_{1,t} = g_{2,t} \).\\
\emph{Action} (a): Replace the row with the maximum span-length by the product \( g_{1} g_{2} \).\\
\emph{Condition} (b): Three rows in \( G \), denoted as \( g_1 \), \( g_2 \), and \( g_3 \), have the same left (right) index.\\
\emph{Action} (b): If two of these rows satisfy condition (a), then perform the replacement described in (a). Otherwise, if the components \( g_{1,t} \), \( g_{2,t} \), and \( g_{3,t} \) are different (i.e., they are equal to \( X \), \( Y \), and \( Z \) in some order), replace the row with the maximum span-length by the product \( g_{1} g_{2} g_3 \).

 Algorithm \ref{algLR} terminates because the span length of \(G\) decreases after each iteration. Upon completion, no lines in \(G\) fulfill conditions (a) or (b). Therefore, in accordance with Definition \ref{def:LR}, the resulting matrix \(G\) exhibits the \(LR\) property, i.e., it is in the TOF.

\subsection{Using merging of trellis vertices}

Let \(T\) represent a code trellis for the code \(N\), e.g.  the trivial trellis, where each word in \(N\) corresponds to a unique path. Alternatively, \(T\) could originate from a generator matrix of the code \(N\) without being transformed into LR-form. Then, the minimal code trellis of the rectangular code $N$ can be obtained by merging twin vertices in the trellis (see Algorithm \ref{algMerge}). The algorithm halts as the number of vertices in \(T\) decreases after each step. Upon completion, the output is a trellis without twins and \ES{thus biproper and minimal.}

Other methods, such as Forney or Massey construction, can also be employed to design the minimal trellis. However, all methods yield the same unique minimal trellis.

\begin{algorithm}[t]
    Input: A trellis $T$ for the code $C$.\\
    \quad Find twin vertices in $T$ and merge them.\\
        \quad\quad If twins were not found, output $T$ and stop. 
        \caption{Merging algorithm}\label{algMerge}
\end{algorithm}


\section{Minimal Trellises for Degenerate Decoding}\label{sec:tdegen}

In this section, our focus shifts to reducing the complexity of the DML decoder. \ES{For non-degenerate decoding, our goal was to find the minimal weight path in the trellis representing the code \(\rho N\). For degenerate decoding, our goal is to find the most probable coset \(\ell \rho S\) in \(\rho N\). }

This requirement suggests that {for DML trellis decoding each coset of $S$ in $\rho N$, that is, all $\ell S$ for all $\ell \in L$}, should have its dedicated goal node, where all paths representing that coset terminate. Subsequently, the {Viterbi} sum-product algorithm can be employed to compute the total probability of the paths terminating at each node. The coset with the highest probability is then selected.

Alternatively, one can view the code \(N\) as a collection of subcodes \(\ell S\), where \(\ell \in L\), \ES{called \emph{joint} code introduce here}.  {In order to design the minimal trellis representing all cosets of $S$}, we must define ``multi-goal'' trellises and introduce some of their properties.

\subsection{Multi-goal trellises and Joint Code}
In this section, we introduce the multi-goal trellises for block codes and show some of their properties. A similar definition of multi-goal-trellises were proposed in \cite{li2019maximum}.


Consider Definition~\ref{def:trellis}, we still assume that the \emph{subset $\mathcal{V}_0$ consists of a single vertex.} \ES{However, there can be multiple \emph{goal}} vertices  $v_{g}\in \mathcal{V}_n$. To emphasize that, when $|\mathcal{V}_n|=m> 1$, we can call the trellis an \emph{$m$-multi-goal-trellis}. Otherwise it can be called a \emph{simple trellis}.

An example of a 4-multi-goal-trellis of depth $4$ with the alphabet $A=\{I,X,Y,Z\}$ is shown in Figure~\ref{fig:BCJR} which is the trellis of all the cosets of $S$ in $P_1^n$ for the 4-qubit code.


Let \(T\) be an \(m\)-multi-goal-trellis of depth \(n\) with \(\mathcal{V}_n = \{v_1, \dots, v_m\}\). Here, \(T(v_i)\) represents the simple sub-trellis with a single goal \(v_i\) for \(i = 1, \dots, m\). In other words, \(T(v_i)\) is the sub-trellis of \(T\) that exclusively includes paths terminating at the goal node \(v_i\). Consequently, \(T(v_i)\) presents the code \(C_i\).

\begin{definition}[Joint-code]\label{def:multi-goal-code}
Given the  list of $m$ classical codes $\jc[L]=\{C_1,\dots,C_m\}$. The code $C = \cup_i^m C_i$ together with the list $\jc[L]$ of its subcodes  is called the $m$-joint-code and is denoted by $\jc$.   
\end{definition}

Given an $m$-joint-code $\jc$, we define an \emph{extended joint-code} $C^+$ as a \emph{classical code}  of length $n^+ = n+1$ that contains complete information about \jc, \ES{i.e., the list \jc[L]}. To do this, we take a set of ``tails''
{${\mathfrak T}=\{ \tau_1,\dots,\tau_m \}$}, 
{with distinct $\tau_i$'s}, as the alphabet $A_{n+1}= {\mathfrak T}$ at depth $n+1$ and define:
\begin{equation}\label{ExtCode}
C^+ = \cup_i (C_i,\tau_i).
\end{equation}
For a word $c^+=(c,\tau_i) \in C^+$, the symbol 
$c^+_{n+1} = \tau_i \in \mathfrak T$ shows that the word $c\in C$
belongs to the subcode $C_i$ in the joint-code \jc. \ES{We will later use as $\mathfrak T$ a set of vectors of fixed length.}

We say that the \emph{$m$ multi-goal-trellis $T$ of depth $n$  represents the $m$-joint-code} $\jc$ if each sub-trellis $T(v_i)$ represents the subcode $C_i$ for all $i=1,\dots,m$. 



\begin{definition}[Rectangular joint-code]
A joint-code $\jc[C]$ is called rectangular if the (classical) code $C^+$ is rectangular.
\end{definition}

The minimal multi-goal-trellis for a rectangular $m$-joint-code $\jc$ can be constructed as follows.
The extended code $C^+$ is a classical rectangular code. Design the  minimal trellis $T(C^+)$ of depth $n^+=n+1$ 
using any of the known methods. 
The trellis is biproper and unique by Theorem~\ref{Th2Ksch}. Denote by $T(\jc)$ the sub-trellis of $T(C^+)$ containing the initial $n$ sections. {For multi-goal trellises, we~have:}
\begin{theorem}\label{th:joint-trellis}
	\ES{The trellis} $T(\jc)$ is a unique minimal biproper $m$-multi-goal trellis of the $m$-joint-code $\jc$~minimizing
		\begin{enumerate}
		\item $|\cV|$, total number of vertices; 
		\item $|\cE|$, total number of edges;
		\item $|\cE|-|\cV|$, the cyclic rank of $T$. 
	\end{enumerate}
\end{theorem}
\begin{proof}
	Denote  $T(\jc)= (\cV,\cE)$ and   $T(C^+)= (\cV^+,\cE^+)$. \ES{In the $n$-th level, we have that} $\cV_n = \cV^+_{n}$ and $|\cV_n|= m$ since the $m$ tails $\tau_i$ are different and  $T(C^+)$ is minimal and biproper. 
 Since $T(\jc)$  was obtained by deleting the last $(n+1)$-st section from $T(C^+)$, $T(\jc)$ is biproper and we have the following relations:
   \begin{align*}
     	  |\cV|&=|\cV^+|-1,\\
	 |\cE|&=|\cE^+| - m,\\
	  |\cE| - |\cV| &=|\cE^+| - |\cV^+|-m+1.  
   \end{align*}
	  Hence minimization of  $T(C^+)$ simultaneously minimizes $T(\jc)$, and the theorem statement follows.
	  \end{proof}
Analogously to Theorem~\ref{Th2Ksch}, we have:

  \begin{corollary}\label{cor1}
  	For a joint-code $\jc$ the following are equivalent:
  \begin{enumerate}
      \item  \jc \  has a reduced biproper multi-goal trellis representation $T$;
      \item $\jc$ is rectangular;
      \item $T$ is a unique minimal multi-goal trellis for $\jc$.
  \end{enumerate} 
  \end{corollary}	
  
{Therefore, according to Theorem~\ref{th:joint-trellis} and Corollary~\ref{cor1}, ensuring the trellis is biproper is sufficient for obtaining the minimal multi-goal trellis $T$ for a rectangular joint code $\jc$.}

{We are now ready to express the cosets of $S$ in $N$ as a joint code defined in Definition~\ref{def:multi-goal-code}. Since the stabilizer group $S$ is a subgroup of the normalizer $N$ we can partition $N$ into disjoint cosets $N=\cup_{\ell\in L} N_\ell$, where $N_{\ell} = \ell S$ and $\ell$ are representatives of the cosets. The representatives $\ell$ form the factor group $L=N/S$, $|L|=2^{2k}$, of logical operators, generated by the matrix $G(L)\in P_1^{2k\times n}$,  which has as rows the generators of the group $L$.
The code $N$ with the list of subcodes $N_\ell$, which are cosets of $S$ in $N$,   
\begin{equation*}\label{eq:bS}
\jc[L]=\{N_\ell: \ell \in L\}=\{\ell S: \ell \in L\},
\end{equation*} 
is the joint-code $\jc[N]$, also denoted by the pair $(N,S)$. 
In general, let $C$ be a group code and $C'$ be its subcode. Then $(C,C')$ denotes the joint code $\jc$ where  the list $\jc[L]$ consists of the cosets of $C'$ in $C$.}


\subsection{Shannon product approach}
\label{sec:Shan}
{In this section, we use the Shannon (trellis) product to construct the minimal multi-goal trellis for the joint code $\jc[N]$. First, consider the following example.}
\begin{example}
The generator matrix $G(N) \in P_1^{(n+k) \times n}$ with the generators of N as rows of the $[[4,2]]$ code \cite{vaidman1996error} is given by
\begin{align}
G(N)= \begin{pmatrix}
   X&X&X&X\\
Z&Z&Z&Z\\
\hline
I&X&X&I\\
I&Z&Z&I\\
I&I&X&X\\
I&I&Z&Z\\ 
\end{pmatrix} = \begin{pmatrix}
G(S)\\
\hline
G(\cs)\end{pmatrix} ,\label{eq:four}
\end{align}
where $G(S)\in P_1^{(n-k)\times n}$ has as rows the generators of the stabilizer group.
\end{example}
The extended code $N^+$ like in \eqref{ExtCode}
for the $[[4,2]]$ code of Example~1 can be generated by the extension of the generator matrix \eqref{eq:four} as follows:
\begin{align}
    G^+ = \begin{pmatrix}
X&X&X&X&\vline& I&I&I&I\\
Z&Z&Z&Z&\vline&I&I&I&I\\
I&X&X&I&\vline&X&I&I&I\\
I&Z&Z&I&\vline&I&X&I&I\\
I&I&X&X&\vline&I&I&X&I\\
I&I&Z&Z&\vline&I&I&I&X\\    
\end{pmatrix} \label{eq:gen_joint1},
\end{align}
{where we used a set of tails $\jc[T]$ of $4$-vectors over $P_1$. {These vectors  $(IIII)$, $(XIII)$,..., $(IIIX)$ generate the set of tails $\jc[T]$, where every element of $\jc[T]$ can be considered as a symbol in the alphabet $A_{n+1} = \jc[T]$. 
Then $G$ generates the (two-alphabetic) code of length $n+1$ like in Definition~\ref{def:multi-goal-code}}. In general, we have the following Lemma.}
\begin{lemma}\label{eq:G(C+)}
Given the joint-code $\jc[N]$, the matrix,
    \begin{align}
        G^+ = \begin{pmatrix}
        G(S^+) \\
        G(L^+)
    \end{pmatrix} =\begin{pmatrix}
        G(S) & Q_1 \\
        G(L) & Q_2
    \end{pmatrix} \in P_1^{(n+k)\times (n+2k)}\label{extension},
    \end{align}
where the rows of $Q_1$ are the all identity, i.e., $(I,\dots,I)$ and the $j_{th}$ row of $Q_2$  has $X$ in the $j_{\text{th}}$ position and identities elsewhere, is a generator matrix of the extension code $C^+$. The code $C^+$ is a rectangular code of length $n+2k$ (equivalent to length $n+1$ for the two-alphabetic code) and dimension $n+k$.
\end{lemma}
\begin{proof} 
	The code $N^+$ generated by the $(n+k)\times(n+2k)$-matrix $G^+$ is a group code and, hence, it is a rectangular code~\cite{kschischang1995trellis}.
	To prove the first statement, we should show that the tails {$\tau_{i}$} in  \eqref{extension}
	 are different for different cosets. This is true since every codeword of $N^+$  has form $\ell s$, for 
	  $\ell\in L^+$ and $s\in S^+$, and the selected tails of $L^+$ are independent. 
	\end{proof} 
 
Note that the selection of tails to design the extended matrix is not unique. The only requirement is that the tails of different cosets of $S$ should be distinct. Another example of an extension matrix with quaternary tails of length $k$ is given by the rows of $Q_2$
that are equal to $\{(X,I),(Z,I),(I,X),(I,Z)\}$ and generate the set of the tails $\jc[T]$.

 The minimal trellis of the rectangular code $C^+$, which is unique by Theorem~\ref{Th2Ksch}, can be obtained using the matrix $G^+$ in \eqref{eq:gen_joint1} in the {TOF} or $G(N)$ in the restricted TOF defined below.
\begin{figure}[t]
    \centering




\tikzstyle{state}=[shape=circle,draw=blue!50,fill=blue!20, inner sep=1pt,minimum size=1pt]
\tikzstyle{observation}=[shape=rectangle,draw=orange!50,fill=orange!20]
\tikzstyle{lightedge}=[<-,dotted]
\tikzstyle{mainstate}=[state,thick]
\tikzstyle{mainedge}=[<-,thick]
\tikzstyle{identity}=[<-,solid]
\tikzstyle{Xop}=[<-,dashed, color = red]
\tikzstyle{Zop}=[<-,densely dotted, color = blue]
\tikzstyle{Yop}=[<-,dashdotted, color = green]

\begin{tikzpicture}[]
\node[state] (s1_1_3) at (0,5) {\tiny $0$};
\node[state] (s1_2_3) at (2,5) {\tiny$0$}
    edge[identity] (s1_1_3);
\node[state] (s1_3_3) at (4,5) {\tiny$0$}
    edge[identity]  (s1_2_3);

\node[state] (s2_3_3) at (4,4.5) {\tiny$1$}
    edge[Zop] (s1_2_3);    

\node[state] (s1_4_3) at (6,5) {\tiny$0$}
    edge[identity]  (s1_3_3);
    
\node[state] (s2_4_3) at (6,4.5) {\tiny$0$}
    edge[Zop] (s2_3_3);
    
\node[state] (s1_5_3) at (8,5) {\tiny $0$}
    edge[identity]  (s1_4_3);

\node[state] (s2_5_3) at (8,4.5) {\tiny $1$}
    edge[identity]  (s2_4_3);



    
    


\node[state] (s1_1_4) at (0,4) {\tiny $0$};
\node[state] (s1_2_4) at (2,4) {\tiny$0$}
    edge[identity] (s1_1_4);
\node[state] (s1_3_4) at (4,4) {\tiny$0$}
    edge[identity]  (s1_2_4);

\node[state] (s1_4_4) at (6,4) {\tiny$0$}
    edge[identity]  (s1_3_4);
    
\node[state] (s2_4_4) at (6,3.5) {\tiny$0$}
    edge[Xop] (s1_3_4);
    
\node[state] (s1_5_4) at (8,4) {\tiny $0$}
    edge[identity]  (s1_4_4);

\node[state] (s2_5_4) at (8,3.5) {\tiny $1$}
    edge[Xop]  (s2_4_4);


    
    

    
\end{tikzpicture}
        \caption{{Multi-goal atomic trellises of $l^{(1)} = IXXI$ and $l^{(2)} = IIZZ$.}}
        \label{fig:gen_2}
\end{figure}
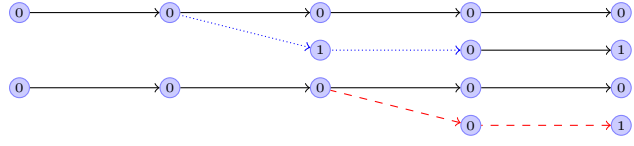

\begin{definition}[Restricted TOF] \label{def:restof} We say that the matrix $G(N)$ like in \eqref{eq:four} is in the restricted TOF if it has the L-property and the submatrix $G(S)$ is in the TOF.
\end{definition}
{After the extension of $G(N)$, which is in the restricted TOF, described in Lemma~\ref{eq:G(C+)}, $G^+$ will be in the TOF, e.g., see \eqref{eq:gen_joint1}.} 

To design the minimal multi-goal-trellis for the joint code~$\jc[N]$, one can proceed as follows.

(1) Compute the Shannon product of atomic trellises \cite{kschischang1995trellis,Sidorenko1996} for $G^+$ in TOF to obtain the minimal trellis $T^+$ of the code $N^+$.

(2)  By Theorem~\ref{th:joint-trellis}, the first $n$ sections of $T^+$ give us the unique minimal multi-goal-trellis for the joint code $\jc[N]$. 

We can directly compute the state profile $|\cV_t|$ and edge profile $|\cE_t|$ of the multi-goal trellis from $G$ using known classical methods \cite{mceliece1996bcjr}. For \eqref{eq:gen_joint1}, the state profile is 1, 4, 16, 64, 16, and the edge profile is 4, 16, 64, 64.

\subsection{Shannon product of atomic multi-goal trellises} \label{sec:shannonmult}
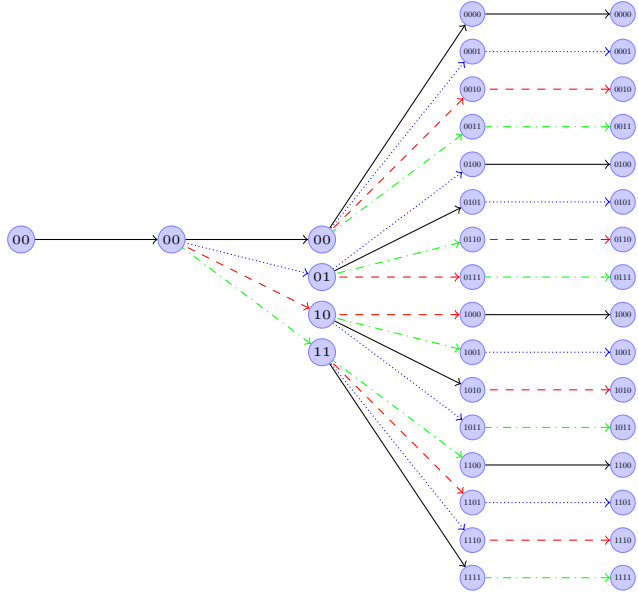
\begin{figure}[t]
    \centering
    {




\tikzstyle{state}=[shape=circle,draw=blue!50,fill=blue!20, inner sep=1pt,minimum size=1pt]
\tikzstyle{observation}=[shape=rectangle,draw=orange!50,fill=orange!20]
\tikzstyle{lightedge}=[<-,dotted]
\tikzstyle{mainstate}=[state,thick]
\tikzstyle{mainedge}=[<-,thick]
\tikzstyle{identity}=[<-,solid]
\tikzstyle{Xop}=[<-,dashed, color = red]
\tikzstyle{Zop}=[<-,densely dotted, color = blue]
\tikzstyle{Yop}=[<-,dashdotted, color = green]
\begin{tikzpicture}[]
\node[state] (s1_1) at (0,5) {\tiny $00$};
\node[state] (s1_2) at (2,5) {\tiny$00$}
    edge[identity] (s1_1);

\node[state] (s1_3) at (4,5) {\tiny$00$}
    edge[identity]  (s1_2);
\node[state] (s2_3) at (4,4.5) {\tiny$01$}
    edge[Zop] (s1_2);
\node[state] (s3_3) at (4,4) {\tiny$10$}
    edge[Xop] (s1_2);
\node[state] (s4_3) at (4,3.5) {\tiny$11$}
    edge[Yop] (s1_2);

\node[state] (s1_4) at (6,8) {\scalebox{.35}{$0000$}}
    edge[identity]  (s1_3);
\node[state] (s2_4) at (6,7.5) {\scalebox{.35}{$0001$}}
    edge[Zop] (s1_3);
\node[state] (s3_4) at (6,7) {\scalebox{.35}{$0010$}}
    edge[Xop] (s1_3);

\node[state] (s4_4) at (6,6.5) {\scalebox{.35}{$0011$}}
    edge[Yop] (s1_3);
    
\node[state] (s5_4) at (6,6) {\scalebox{.35}{$0100$}}
    edge[Zop]  (s2_3);
\node[state] (s6_4) at (6,5.5) {\scalebox{.35}{$0101$}}
    edge[identity] (s2_3);
\node[state] (s7_4) at (6,5) {\scalebox{.35}{$0110$}}
    edge[Yop] (s2_3);

\node[state] (s8_4) at (6,4.5) {\scalebox{.35}{$0111$}}
    edge[Xop] (s2_3);

\node[state] (s9_4) at (6,4) {\scalebox{.35}{$1000$}}
    edge[Xop]  (s3_3);
\node[state] (s10_4) at (6,3.5) {\scalebox{.35}{$1001$}}
    edge[Yop] (s3_3);
\node[state] (s11_4) at (6,3) {\scalebox{.35}{$1010$}}
    edge[identity] (s3_3);

\node[state] (s12_4) at (6,2.5) {\scalebox{.35}{$1011$}}
    edge[Zop] (s3_3);
    
\node[state] (s13_4) at (6,2) {\scalebox{.35}{$1100$}}
    edge[Yop]  (s4_3);
\node[state] (s14_4) at (6,1.5) {\scalebox{.35}{$1101$}}
    edge[Xop] (s4_3);
\node[state] (s15_4) at (6,1) {\scalebox{.35}{$1110$}}
    edge[Zop] (s4_3);

\node[state] (s16_4) at (6,0.5) {\scalebox{.35}{$1111$}}
    edge[identity] (s4_3);    
    
\node[state] (s1_5) at (8,8) { \scalebox{.35}{$0000$}}
    edge[identity]  (s1_4);

\node[state] (s1_5) at (8,7.5) {\scalebox{.35}{$0001$}}
    edge[Zop] (s2_4);

\node[state] (s1_5) at (8,7) {\scalebox{.35}{$0010$}}
    edge[Xop] (s3_4);

\node[state] (s1_5) at (8,6.5){\scalebox{.35}{$0011$}}
    edge[Yop] (s4_4);

\node[state] (s1_5) at (8,6) { \scalebox{.35}{$0100$}}
    edge[identity]  (s5_4);

\node[state] (s1_5) at (8,5.5) {\scalebox{.35}{$0101$}}
    edge[Zop] (s6_4);

\node[state] (s1_5) at (8,5) {\scalebox{.35}{$0110$}}
    edge[Xop] (s7_4);

\node[state] (s1_5) at (8,4.5){\scalebox{.35}{$0111$}}
    edge[Yop] (s8_4);

\node[state] (s1_5) at (8,4) { \scalebox{.35}{$1000$}}
    edge[identity]  (s9_4);

\node[state] (s1_5) at (8,3.5) {\scalebox{.35}{$1001$}}
    edge[Zop] (s10_4);

\node[state] (s1_5) at (8,3) {\scalebox{.35}{$1010$}}
    edge[Xop] (s11_4);

\node[state] (s1_5) at (8,2.5){\scalebox{.35}{$1011$}}
    edge[Yop] (s12_4);

\node[state] (s1_5) at (8,2) { \scalebox{.35}{$1100$}}
    edge[identity]  (s13_4);

\node[state] (s1_5) at (8,1.5) {\scalebox{.35}{$1101$}}
    edge[Zop] (s14_4);

\node[state] (s1_5) at (8,1) {\scalebox{.35}{$1110$}}
    edge[Xop] (s15_4);

\node[state] (s1_5) at (8,0.5){\scalebox{.35}{$1111$}}
    edge[Yop] (s16_4);

\end{tikzpicture}

}
        \caption{{Multi-goal trellis of $L$}.}
        \label{fig:genshann}
\end{figure}

This section acts as an extension of the previous one. 
To construct the minimal multi-goal trellis,
we bring $G(N)$ into the restricted TOF of Definition~\ref{def:restof}. Then, the minimal multi-goal trellis can be obtained by the Shannon product of ``multi-goal atomic trellises'' defined next,

\begin{definition}[Multi-goal atomic trellis] \label{def:ma}
{Denote by $T(l)$, where $l\in P_1^n$, the minimal trellis of the code {$\{(I,\dots,I), l\} $} with two paths ending at two different goal nodes: the $l$ path and the all identity $(I,\dots, I)$ path.}
\end{definition}
\ES{Note that this definition only differs from the classical one \cite{kschischang1995trellis}, in the constraint that the two paths should end in a single goal node.}
For an example of multi-goal atomic trellises of the first two rows in \eqref{eq:four}, see Figure \ref{fig:gen_2} where {the top multi-goal trellis, i.e.,  $T((I,X,X,I))$ has two paths, one for $(I,I,I,I)$ and one for $(I,X,X,I)$ both ending in distinct nodes}. {By using $G(L)$, one can compute the trellis $T_L$ of $L$ by taking the Shannon product of the multi-goal atomic trellises of all $l^{(i)}$, i.e., the rows of $G(L)$. For the matrix in \eqref{eq:four}, it results in Figure~\ref{fig:genshann}. {It is easy to verify that the paths of the trellis in Figure~\ref{fig:genshann} are exactly the $2^{2k}$ logical operators in $L$.}} Finally, we compute the Shannon product of $T_L$ and the minimal trellis of the stabilizer group $T_S$ and obtain the minimal multi-goal trellis of the code.
\begin{corollary}
Given a matrix $G(N)$ in the restricted TOF, the trellis $T$ obtained by the Shannon product of the multi-goal atomic trellises $T(l^{(i)})$ of $G(L)$ and the minimal trellis $T_S$ of the stabilizer group is the minimal multi-goal trellis of the joint code $\jc[N]$.
\end{corollary} 
\begin{proof}  The Shannon product of the multi-goal atomic trellises will generate $2^{2k}$ goal nodes, one for every logical operator. {By then taking the Shannon product with the minimal trellis of the stabilizer group, we present all cosets of $S$ in $N$. The resulting trellis is biproper and, by Corollary~1, minimal.}
\end{proof}
 {Note that, with the above construction, we create the minimal multi-goal trellis of Section~\ref{sec:Shan} by computing the product of the first $n$-sections of the atomic trellises of $G^+$.}

\subsection{Using the BCJR-Wolf trellis}\label{sec:Wolf}

Consider a group code $U$ of length $n$ over the alphabet $P_1$ with a check  matrix $H$ of size $r\times n$. The space $W=P_1^n$ can be partitioned into $m=2^{r}$ cosets of the code $U$,  $W = \cup_{b\in W}\,  bU$, and we obtain
 a joint code $\jc[W]= (W,U)$. Using the BCJR-Wolf method, defined in Section~\ref{sec:wolfnorm}, we can derive the minimal $m$-multi-goal-trellis $T$ representing $\jc[W]$. The goal vertices $\mathcal{V}_n$ in $T$ are labeled by the syndromes of the cosets, defined as $\sigma(b) = b * H^T$, where $b$ is a representative of the coset. The trellis $T$ is also called the complete trellis for the code $U$.
For example, Figure 1 illustrates the complete trellis for the normalizer code $N$ of the $[[4,2]]$ code, which is the $m$-multi-goal-trellis, $m=4$, of the joint code $(P_1^n,N)$. Here, the matrix $ G(S) = H$ is defined in \eqref{eq:four} with parameters $r=2$ and $m=4$.

For degenerate decoding, we need to design the minimal multi-goal-trellis $T$ for the joint code ${(N,S)}$.
Using the BCJR-Wolf algorithm with check matrix $H = G(N)$, we construct the $m$-multi-goal-trellis $T'$ for $(P_1^n,S)$ with $m=2^{n+k}=|\cV_n|$. Due to the BCJR-Wolf algorithm, the trellis $T'$ is biproper \cite{wolf1978efficient}. To form the minimal trellis of $(N,S)$, we retain only the cosets $bS$ of $S$ within $N$. The initial $n-k$ rows of $G(N)$, constituting $G(S)$, ensure that codewords of $N$ exhibit zero syndromes in these positions. Therefore, we selectively preserve cosets with zero syndromes in the initial $n-k$ positions.
After removing unnecessary vertices from $\cV_n$ we obtain the required $m$-multi-goal-trellis $T$ with $m=2^{2k}$ goal vertices. The trellis $T$ is still biproper and hence, by Corollary~\ref{cor1}, minimal.
In the BCJR-Wolf approach, the vertices $\cV_t$ at any depth $t$ are indexed by binary vectors of length $r=n+k$; hence we have the BCJR-Wolf bound for the state complexity of $T$,
\begin{equation}
|\cV_t|\le 2^{n+k} \quad \forall t. \label{Wolf_bound}
\end{equation} 
In the case of stabilizer codes, the bound also follows from the fact that $|N|=  2^{n+k}$.

\subsection{Merged trellises of cosets of $S$}
\label{sec:merge}

{For each of the $2^{2k}$ cosets $\ell S$ with $\ell\in L,$ we build their minimal trellises. These trellises can be constructed by relabeling the edges of the minimal trellis of the stabilizer group, based on the corresponding element $\ell_i$, $i=1,\dots,n$, of the representative $\ell=(\ell_1,\dots,\ell_n)$. {In \cite{sabo2022trellis}, the authors used those trellises for degenerate decoding; however, here we use these trellises to build the \emph{minimal} multi-goal trellis.} To do this we {merge} the root vertices of these trellises to form a non-minimal multi-goal trellis $T'.$ Since the joint code $C$ is rectangular, we merge twin vertices in $T'$ until obtaining the biproper multi-goal trellis $T.$ \ES{Corollary~\ref{cor1}, then implies that the trellis $T$ is the unique minimal trellis for the joint code $C$.}}


\section{Minimal Trellis for Degenerate Decoding of CSS codes for Independent Errors} \label{sec:dmlcss}
In this section, we consider CSS codes \cite{calderbank1996good, calderbank1998quantum}. One key characteristic of CSS codes is the ability to partition the generators of the stabilizer group \(S\) and the normalizer group into purely \(X\)-generators and purely \(Z\)-generators.

Before we recall CSS codes we introduce some notation. {By $X^\alpha$ where $\alpha \in \{0,1\}^n$} 
we denote the vector, $(X^{\alpha_1}, X^{\alpha_2}, \cdots,  X^{\alpha_n})\in P_1^n$ where $X^0 = I$ (the identity) and $X^1 = X$. \ES{For example, if $\alpha = (0, 1, 1)$ then $X^{\alpha} = (I,X,X)$. We use a similar notation for $Z^\alpha$.}
We denote a classical code $C$ of length $n$, dimension $k$ and minimum distance $d$ by an  $[n,k,d]$. By $C^\perp$ we denote the dual of that code which comprises of all vectors \(v \in \mathbb{F}_2^{n}\) such that \(v\cdot u = 0\) for all \(u \in C\), where `\(\cdot\)' denotes the usual inner product between vectors. Define the alphabets $A_X = \{I,X\}$ and $A_Z = \{ I,Z\}$ and denote the space of $n$-vectors over those alphabets by $A_X^n$ and $A_Z^n$ respectively.
\ES{We now recall the definition of CSS codes.}

\begin{definition}[CSS codes] Given two classical binary codes $C_1,C_2 \subseteq \mathbb{F}_2^n$ with parameters \([n,k_1,d_1]\) and \([n,k_2,d_2]\) respectively, such that \(C_2^\perp \subseteq C_1\) the associated CSS code has stabilizer group $S = \langle S_X,S_Z \rangle $ where $S_X = \{X^{\alpha}:\: \alpha \in C_1^\perp\}$ and $S_Z = \{Z^{\beta}:\: \beta \in C_2^\perp\}$ and it is an $[[n,k_1 +
 k_2-n]]$ stabilizer code.
\end{definition}

\ES{Let $H_1,H_2$ be the parity check matrices of $C_1$ and $C_2$ respectively} and denote their rows by $h_j^{(i)},i=1,\dots, n-k_j,j=1,2$. Then the $X$- and $Z$-stabilizers, i.e., $S_X$ and $S_Z$ respectively, can be generated by,
\begin{align}
    &S_X = \langle s^i_X = X^{h_1^{(i)}}:\: i=1,\dots, n-k_1 \rangle,\\ &S_Z = \langle s^i_Z =  Z^{h_2^{(i)}}:\: i=1,\dots, n-k_2 \rangle \nonumber .
\end{align} 
\ES{Recall that, for an $[[n, k]]$ stabilizer code,} the syndrome $\sigma$ of an error $e$ is computed by $\sigma_i = e * s_i$ for $i=1, \dots, n-k$. Any error $e \in P_1^n$ can be decomposed as $e = e_X e_Z$, where $e_X \in A_X^n$ and $e_Z \in A_Z^n$. Since we can split the stabilizers into purely $X$- and purely $Z$-stabilizers, the syndrome can also be decomposed into two parts. This is because the $e_X$ part of the error $e$ always commutes with the $X$-stabilizers, which results in a zero syndrome, and similarly for the $e_Z$ part of the error $e$ and the $Z$-stabilizers. By $\sigma_X\in \mathbb{F}_2^{n-k_1}$ and $\sigma_Z\in \mathbb{F}_2^{n-k_2}$ we denote the syndrome obtained by the $X$- and $Z$-stabilizers, that is, 
\begin{align*}
 \sigma_{X,i} =  e_Z * s_X^i,\;i=1,\dots, n-k_1,\quad \sigma_{Z,i} = e_X * s_Z^i,\; i= 1,\dots, n-k_2 .
\end{align*}
Moreover, the division of the stabilizers also allows for the division of the \(k_1+k_2\) generators of the normalizer code \(N\) into $k_2$ purely \(X\)- and $k_1$ purely \(Z\)-generators \cite{wilde2009logical} generating the $N_X$ over the binary alphabet $A_X$ and $N_Z$ over the alphabet $A_Z$ respectively. Clearly, the full code {over the quaternary alphabet $P_1$} is given by $N = N_X \times N_Z$ as defined in~\eqref{product}.

The authors in \cite{sabo2022trellis} illustrate how one can perform independent decoding of the $X$ and $Z$ stabilizers using trellises. This is achieved by constructing {separate trellises} for the binary code $N_X$ and for the binary code $N_Z$. In the following, we extend this concept to the DML decoding of the binary codes. Note that if the $X$ and $Z$ errors are independent, then the separate decoding of $N_X$ and $N_Z$ is an NDML decoding technique, which is one of the prominent methods of decoding CSS codes\cite{demarti2023decoding}. However, if the $X$ and $Z$ errors are not independent, e.g., in the depolarizing channel, then this decoding approach is not an NDML decoding. Nonetheless, such decoders perform well, e.g., the MWPM decoder \cite{fowler2013minimum} and the Union-find decoder~\cite{delfosse2021almost}. 

\subsection{Degenerate decoding of CSS codes}
Here, we investigate the separate DML decoding of CSS codes. Note that the code $N_Z$ consists of all the operators in $A_Z^n$ that commute with the stabilizers in $S_X$ and the code $N_X$ consists of all the operators in $ A_X^n$ that commute with the stabilizers in $S_Z$.

Let us first examine the binary code $N_Z$ alongside the stabilizer generators $S_X$. Suppose we measure $\sigma_X(e_Z)$ where $e_Z \in A_Z^n$. By definition, $\sigma_X(e_Z) = \sigma_X(b e_Z)$ for all $b \in N_Z$. Consequently, the factor group $A_Z^n /N_Z$ comprises $2^{n-k_1}$ cosets, where the words within a coset share the same syndrome. We denote the representatives of these cosets as $\rho_Z\in A_Z^n$. However, similar to the general case, any error vectors $e,e'\in A_Z^n$ that exert the same effect on the code satisfy $e = e's_Z $ for some $s_Z\in S_Z$. Since both errors have identical syndromes $\sigma_X$ with a representative $\rho_Z$, but vary by a stabilizer element, they belong to one of the cosets of the stabilizer $S_Z$ in the normalizer coset $\rho_Z N_Z$. Given $|N_Z| = 2^{k_1}$ and $|S_Z|=2^{n-k_1}$, this implies the existence of $2^{k_1 + k_2 - n} = 2^{k}$ such cosets. We label the group of representatives of these cosets as $L_Z$ and its generators as $\{\ell_Z^i\}_{i=1}^{{k}},\:\ell_i \in A_Z^n$,  referring to its elements as $Z$ logical operators.

Similarly, considering the binary code $N_X$, the factor group $A_X^n /N_X$ comprises $2^{n-k_2}$ cosets, where the words within a coset share the same syndrome. We denote the representatives of these cosets as $\rho_X\in A_X^n$. Furthermore, with analogous arguments as before, we group and denote the representatives of the $2^{k}$ cosets of $S_X$ in $\rho_X N_X$ as $L_X$, with its generators labeled as $\{\ell_X^i\}_{i=1}^{{k}},\:\ell_i \in A_X^n$, and we refer to its elements as $X$ logical operators.

Assuming that $X$ and $Z$ errors are independent, then similarly as in the previous sections, the DML decoder should choose \emph{independently} the coset of $S_X$ in $N_X$ and the coset of $S_Z$ in $N_Z$  with the largest probability. By the above factorization, any error $e\in P^n_1$ can be decomposed as $e = (e_X)(e_Z) = (\rho_X \ell_X s_X)(\rho_Z \ell_Z s_Z) $ where $e_X\in A_X^n$, $e_Z\in A_Z^n$. Hence, one can independently find the optimal $\ell_X$ and $\ell_Z$. 

The DML decoding for independent $X$ and $Z$ errors for CSS codes is then summarized as follows: \\ 
(1)  Given the syndromes $\sigma_X$ and $\sigma_Z$, find $\rho_X$ and $\rho_Z$ such that the cosets $\rho_X N_X$ and $\rho_Z N_Z$  have syndromes $\sigma_X$ and $\sigma_Z$, respectively.\\ 
{(2) Compute probabilities  $\text{Pr}(\ell_X)$ and $\text{Pr}(\ell_Z)$ of each coset $\rho_X \ell_X S_X$ {in $\rho_X N_X$} for all $\ell_X \in L_X$ and $\rho_Z \ell_Z S_Z$ {in $\rho_Z N_Z$} for all $\ell_Z \in L_Z$, respectively, using following \emph{sum-product} formulas.}
  \begin{align*}
&\text{Pr}(\ell_X)= \sum_{w\in \ell_X S_X} \prod_{i=1}^n \text{Pr}(\rho_{X,i} w_i),\\
&\text{Pr}(\ell_Z)= \sum_{w\in \ell_Z S_Z} \prod_{i=1}^n \text{Pr}(\rho_{Z,i} w_i).
\end{align*}
(3) Output $\ell = \ell_X \ell_Z$ where $\ell_X$ and $\ell_Z$ have the maximum probability~$\text{Pr}(\ell_X)$ and $\text{Pr}(\ell_Z)$, respectively.\\

To reduce the complexity of Step 2, we use the minimal multi-goal trellises that represent all cosets of $S_X$ in $N_X$ in one trellis and the cosets of $S_Z$ in $N_Z$ in another trellis. We can build these trellises by using all the methods discussed in Section~\ref{sec:tdegen} for each of these trellises individually. 

\subsection{Minimal Trellis construction}

In this section, we expand upon the methods outlined in Section~\ref{sec:tdegen} for the separate DML decoding of CSS codes. The joint codes under consideration are denoted as $\jc[N]_X$ and $\jc[N]_Z$, represented by the pairs $(N_X,S_X)$ and $(N_Z,S_Z)$, respectively. Here, we demonstrate the construction of minimal multi-goal trellises $T_X = T(\jc[N]_Z)$ and $T_Z = T(\jc[N]_X)$. It is important to note that $T_X$, referred to as the $X$-stabilizer multi-goal trellis, contains all the vectors in $A_Z^n$ that commute with the stabilizer $S_X$, i.e., $N_Z$. Thus, we use the subscript $X$ on $T_X$ since with $S_X$, we derive the syndrome. Similarly, $T_Z$ is the $Z$-stabilizer multi-goal trellis.

Let us consider the matrices $G(N_X)$ and $G(N_Z)$, which contain the $X$- and $Z$-generators of $N_X$ and $N_Z$ as rows, respectively. \ES{We can then partition the sub-matrices as follows:}
\begin{align*}
    G(N_X) = \begin{pmatrix}
G(S_X)\\
G(L_X)
\end{pmatrix},\quad    G(N_Z) = \begin{pmatrix}
G(S_Z)\\
G(L_Z)
\end{pmatrix},
\end{align*}
where $G(L_X)\in P_1^{k \times n}$ and $G(L_Z)\in P_1^{k \times n}$ contain rows representing purely $Z$- and $X$-generators of the logical groups $L_Z$ and $L_X$, respectively. 

\begin{example} \label{ex:4} Consider the 7-qubit code $[[7,1,3]]$  \cite{steane1996multiple}. This CSS code is constructed by the $[7,4,3]$ classical Hamming code, which is self-dual, i.e., $C^\perp = C $ which allows $C_2^\perp = C_1$. Therefore, 
\begin{align*}
    H_1 = G_2 = \begin{pmatrix}
         1 & 1 & 1 & 1 & 0 & 0 & 0 \\
         0 & 1 & 1 & 0 & 0 & 1 & 1 \\
         0 & 0 & 1 & 1 & 1 & 1 & 0
    \end{pmatrix}.
\end{align*}
The generators of the code $N_X$ are given by:
\begin{align}
    G(N_X) = \begin{pmatrix}
         X & X & X & X & I & I & I \\
         I & X & X & I & I & X & X \\
         I & I & X & X & X & X & I \\
         \hline 
         I & I & I & I & X & X & X
    \end{pmatrix} = \begin{pmatrix}
        G(S_X)\\
        \hline
        G(L_X)
    \end{pmatrix}. \label{X}
\end{align}
Similarly, the generators of the code $N_Z$ are given by:
\begin{align*}
    G(N_Z) = \begin{pmatrix}
         Z & Z & Z & Z & I & I & I \\
         I & Z & Z & I & I & Z & Z \\
         I & I & Z & Z & Z & Z & I \\
         \hline 
         I & I & I & I & Z & Z & Z
    \end{pmatrix} = \begin{pmatrix}
        G(S_Z)\\
        \hline
        G(L_Z)
    \end{pmatrix}.
\end{align*}
\end{example}
\begin{figure}[t!]
    \centering




\tikzstyle{state}=[shape=circle,draw=blue!50,fill=blue!20, inner sep=1pt,minimum size=1pt]
\tikzstyle{observation}=[shape=rectangle,draw=orange!50,fill=orange!20]
\tikzstyle{lightedge}=[<-,dotted]
\tikzstyle{mainstate}=[state,thick]
\tikzstyle{mainedge}=[<-,thick]
\tikzstyle{identity}=[<-,solid]
\tikzstyle{Xop}=[<-,dashed, color = red]
\tikzstyle{Zop}=[<-,densely dotted, color = blue]
\tikzstyle{Yop}=[<-,dashdotted, color = green]

\begin{tikzpicture}[]
\node[state] (s1_1) at (0.5,5) {\tiny $00$};
\node (I) at (3.5,1) {};
\node[state] (s1_2) at (2,5) {\tiny$00$}
    edge[identity] (s1_1);
\node[state] (s2_2) at (2,4) {\tiny$10$}
    edge[Zop] (s1_1);
\node (I_edge1) at (5.5,1) {$I$} 
edge[identity] (I);
\node (I_edge2) at (6,1) {};
\node[state] (s1_3) at (3.5,5) {\tiny$00$}
    edge[identity]  (s1_2);
\node[state] (s2_3) at (3.5,4.5) {\tiny$01$}
    edge[Zop] (s1_2);
\node[state] (s3_3) at (3.5,4) {\tiny$10$}
    edge[Zop] (s2_2);
\node[state] (s4_3) at (3.5,3.5) {\tiny$11$}
    edge[identity] (s2_2);
\node (X_edge1) at (8,1) {$X$} 
edge[Zop] (I_edge2);
\node[state] (s1_4) at (5,5) {\scalebox{.55}{$000$}}
    edge[identity]  (s1_3);
\node[state] (s2_4) at (5,4.5) {\scalebox{.55}{$001$}}
    edge[Zop] (s1_3);
\node[state] (s3_4) at (5,4) {\scalebox{.55}{$010$}}
     edge[Zop] (s2_3);
\node[state] (s4_4) at (5,3.5) {\scalebox{.55}{$011$}}
    edge[identity] (s2_3);
\node[state] (s5_4) at (5,3) {\scalebox{.55}{$100$}}
    edge[Zop]  (s3_3);
\node[state] (s6_4) at (5,2.5) {\scalebox{.55}{$101$}}
    edge[identity] (s3_3);
\node[state] (s7_4) at (5,2) {\scalebox{.55}{$110$}}
    edge[identity] (s4_3);
\node[state] (s8_4) at (5,1.5) {\scalebox{.55}{$111$}}
    edge[Zop] (s4_3);
\node[state] (s1_5) at (6.5,5) {\tiny$00$}
    edge[identity]  (s1_4)
    edge[Zop]  (s5_4);
\node[state] (s2_5) at (6.5,4.5) {\tiny$01$}
    edge[Zop] (s2_4)
    edge[identity] (s6_4);
\node[state] (s3_5) at (6.5,4) {\tiny$00$}
    edge[identity]  (s3_4)
    edge[Zop] (s7_4);
\node[state] (s4_5) at (6.5,3.5) {\tiny$01$}
    edge[Zop] (s4_4)
    edge[identity] (s8_4);

\node[state] (s1_6) at (8,5) {\scalebox{.55}{$000$}}
    edge[identity]  (s1_5);
\node[state] (s2_6) at (8,4.5) {\scalebox{.55}{$001$}}
    edge[Zop] (s2_5);
\node[state] (s3_6) at (8,4) {\scalebox{.55}{$010$}}
    edge[identity]  (s3_5);
\node[state] (s4_6) at (8,3.5) {\scalebox{.55}{$011$}}
    edge[Zop] (s4_5);

\node[state] (s5_6) at (8,3) {\scalebox{.55}{$100$}}
    edge[Zop]  (s1_5);
\node[state] (s6_6) at (8,2.5) {\scalebox{.55}{$101$}}
    edge[identity] (s2_5);
\node[state] (s7_6) at (8,2) {\scalebox{.55}{$110$}}
    edge[Zop]  (s3_5);
\node[state] (s8_6) at (8,1.5) {\scalebox{.55}{$111$}}
    edge[identity] (s4_5);

\node[state] (s1_7) at (9.5,5) {\tiny$00$}
    edge[identity]  (s1_6)
    edge[Zop]  (s2_6);
\node[state] (s2_7) at (9.5,4.5) {\tiny$01$}
    edge[Zop] (s3_6)
    edge[identity]  (s4_6);

\node[state] (s3_7) at (9.5,3) {\tiny$10$}
    edge[Zop]  (s5_6)
    edge[identity]  (s6_6);
\node[state] (s4_7) at (9.5,2.5) {\tiny$11$}
    edge[identity] (s7_6)
    edge[Zop]  (s8_6);

\node[state] (s1_8) at (11,5) {\tiny$00$}
    edge[identity]  (s1_7)
    edge[Zop]  (s2_7);

\node[state] (s1_8) at (11,3) {\tiny$01$}
    edge[Zop]  (s3_7)
    edge[identity]  (s4_7);
    
\end{tikzpicture}

    \caption{Multi-goal trellis generated by $G(N_X)$ in \eqref{X}.} \vspace{-0.6cm}
    \label{fig:CSS}
\end{figure}
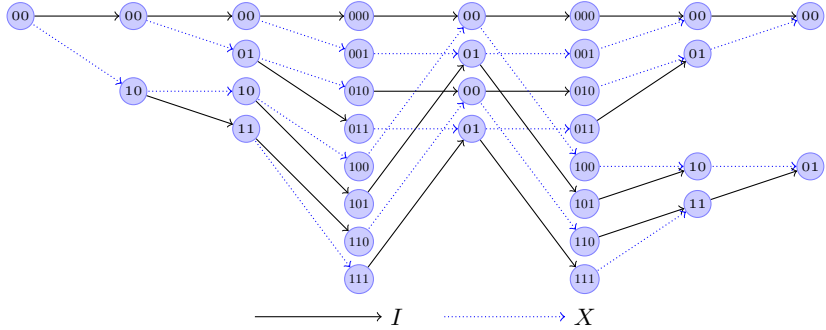

One can then directly apply the construction detailed in Section~\ref{sec:Shan} on \(G(N_X)\) and \(G(N_Z)\) individually to obtain two multi-goal trellises, each of them having $2^{k}$ goal nodes. The $N_X$ multi-goal trellis of the 7-qubit code in Example~\ref{ex:4} can be found in Figure~\ref{fig:CSS}. Similarly, one can apply the construction described in Section~\ref{sec:shannonmult} using multi-goal atomic trellises and the merging algorithm described in Section~\ref{sec:merge} by using \(G(N_X)\) and \(G(N_Z)\) individually. For the BCJR-Wolf method outlined in Section~\ref{sec:Wolf}, instead of constructing the complete trellis of cosets of $S$ in $P_1^n$, \ES{we can build the multi-goal trellis of cosets of $S_X$ in $A_X^n$ and cosets $S_Z$ in $A_Z^n$, respectively, since the codes $N_X$ and $N_Z$ are binary.} 

Consider the following example.

\begin{example}  Consider the $[[7,1,3]]$ code from Example~\ref{ex:4}. Suppose we measure the following syndromes,
\begin{align*}
    \sigma_X = \begin{pmatrix}
        0 & 1 & 0
    \end{pmatrix},\quad \sigma_Z = \begin{pmatrix}
        1 & 0 & 1
    \end{pmatrix}.
\end{align*}
We choose the following vectors as representatives:
\begin{align*}
    &\rho_X = \begin{pmatrix} I& I & X & X & I & I & I 
    \end{pmatrix},\\
    &\rho_Z = \begin{pmatrix} Z& I & I & Z & I & I & I 
    \end{pmatrix}.
\end{align*}
Then we relabel the trellises $T_X$ and $T_Z$ according to $\rho_X$ and $\rho_Z$, respectively, and run the sum-product algorithm {on each trellis separately}. It turns out that the cosets with the largest probabilities are $\rho_X L_X S_X $ and $\rho_Z L_Z S_Z $. We choose an arbitrary error from each coset and form the following joint error:
\begin{align*}
    \hat{e} = \hat{e}_X \hat{e}_Z &= \begin{pmatrix} I& I & I & I & I & I & X 
    \end{pmatrix}\begin{pmatrix} Z& I & Z & I & I & I & Z 
    \end{pmatrix} \\&= \begin{pmatrix} Z& I & Z & I & I & I & Y
    \end{pmatrix}.
\end{align*}
\end{example}

In Table~\ref{table} we compare the reduction of the {decoding} complexity 
compared in building 
for the complete multi-goal trellis $T(\jc[N])$ and the trellises $T_X(\jc[N]_Z
)$ and $T_Z(\jc[N]_X)$. The reduction in complexity is more apparent in the $[[15,1,3]]$ quantum Reed-Muller code \cite{chamberland2017error}.

\begin{table}[t] 
\centering
\begin{tabular}{|l|p{1cm}p{1cm}p{0.5cm}|p{1cm}p{1cm}p{0.5cm}|p{1cm}p{0.5cm}p{0.5cm}|}
\hline
\multirow{2}{*}{}                 & \multicolumn{3}{c|}{$T$}                                                                                                          & \multicolumn{3}{c|}{$T_X$}                                                                                   & \multicolumn{3}{c|}{$T_Z$}                                                                                   \\ \cline{2-10} 
                                  & \multicolumn{1}{p{0.9cm}|}{\centering$|\mathcal{V}|$} & \multicolumn{1}{p{0.9cm}|}{\centering $|\mathcal{E}|$} & \multicolumn{1}{c|}{\centering $2|\mathcal{E}|-|\mathcal{V}|$} & \multicolumn{1}{p{0.9cm}|}{\centering $|\mathcal{V}_X|$} & \multicolumn{1}{p{0.9cm}|}{\centering $|\mathcal{E}_X|$} &\multicolumn{1}{c|}{ \centering $2|\mathcal{E}_X|-|\mathcal{V}_X|$} & \multicolumn{1}{p{0.9cm}|}{\centering $|\mathcal{V}_Z|$} & \multicolumn{1}{p{0.9cm}|}{\centering $|\mathcal{E}_Z|$} & \multicolumn{1}{c|}{ \centering $2|\mathcal{E}_Z|-|\mathcal{V}_Z|$} \\ \hline
\multicolumn{1}{|c|}{$[[4,2,2]]$}  & \multicolumn{1}{c|}{101} & \multicolumn{1}{c|}{148} & \multicolumn{1}{c|}{195} & \multicolumn{1}{c|}{19} & \multicolumn{1}{c|}{22} & \multicolumn{1}{c|}{25} & \multicolumn{1}{c|}{19} & \multicolumn{1}{c|}{22}  & \multicolumn{1}{c|}{25}  \\ \hline
\multicolumn{1}{|c|}{$[[7,1,3]]$} & \multicolumn{1}{c|}{185}  & \multicolumn{1}{c|}{293} & \multicolumn{1}{c|}{401} & \multicolumn{1}{c|}{33} & \multicolumn{1}{c|}{42}  & \multicolumn{1}{c|}{51} & \multicolumn{1}{c|}{33} & \multicolumn{1}{c|}{42}  & \multicolumn{1}{c|}{51}  \\ \hline
\multicolumn{1}{|c|}{$[[9,1,3]]$} & \multicolumn{1}{c|}{81}     & \multicolumn{1}{c|}{131}    & \multicolumn{1}{c|}{183}    & \multicolumn{1}{c|}{27} & \multicolumn{1}{c|}{42} & \multicolumn{1}{c|}{57} & \multicolumn{1}{c|}{27} & \multicolumn{1}{c|}{30} & \multicolumn{1}{c|}{32} \\ \hline
 \multicolumn{1}{|c|}{$[[15,1,3]]$} & \multicolumn{1}{c|}{4773}     & \multicolumn{1}{c|}{8852}    & \multicolumn{1}{c|}{12931}    & \multicolumn{1}{c|}{219} & \multicolumn{1}{c|}{273} &\multicolumn{1}{c|}{246} & \multicolumn{1}{c|}{219} & \multicolumn{1}{c|}{374} & \multicolumn{1}{c|}{529} \\ \hline
\end{tabular}\vspace{0.3cm}
\caption{Trellis complexities for CSS codes for the  multi-goal trellis $T(\jc[N])$, the trellis $T_X(\jc[N]_Z)$ and the trellis $T_Z(\jc[N]_X)$. }\vspace{-0.8cm}
\label{table}
\end{table}

\begin{figure}[htb]
  \centering
  \begin{subfigure}{0.45\textwidth}
    \centering
  \includegraphics[height = 5cm,width=0.95\linewidth]{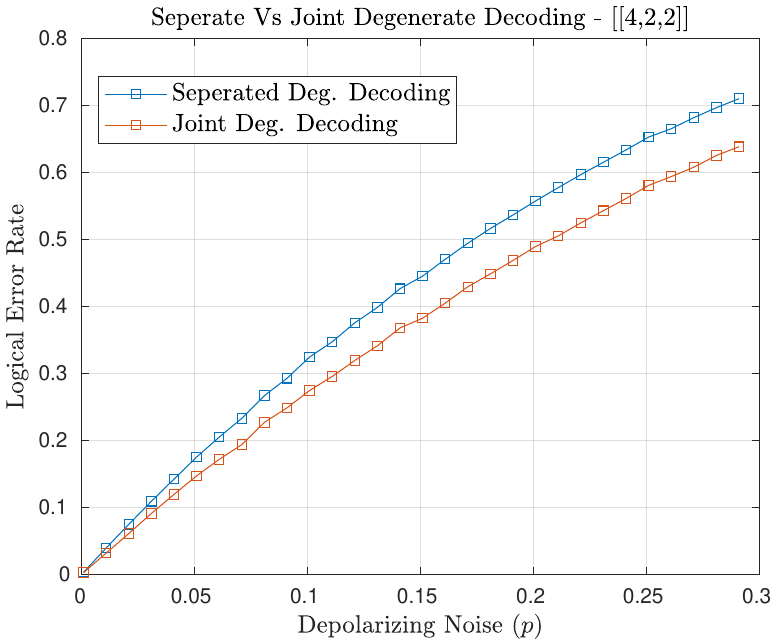}
      \caption{4-qubit code $[[4,2,2]]$}
      \label{fig:4}
  \end{subfigure}
  \hfill                              
  \begin{subfigure}{0.45\textwidth}
    \centering
  \includegraphics[height = 5cm,width=0.95\linewidth]{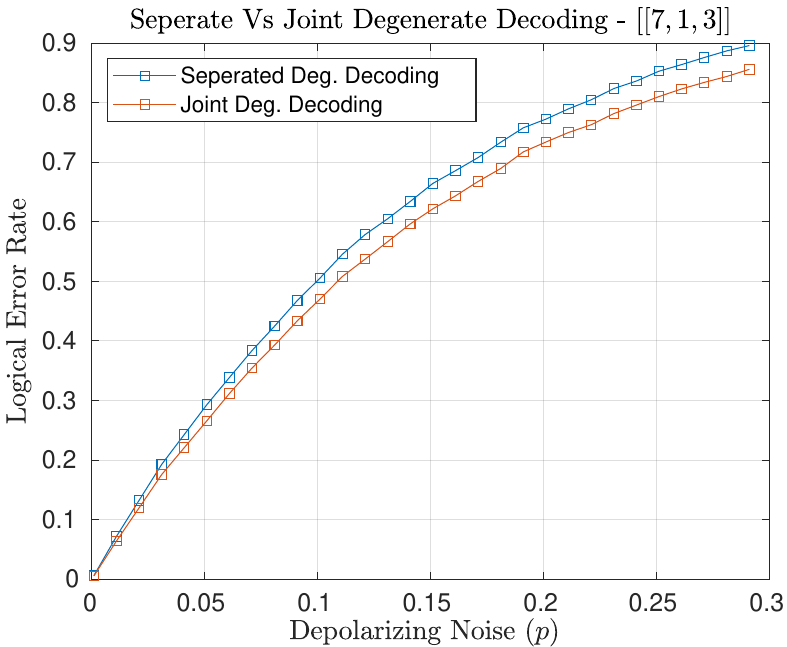}
      \caption{7-qubit code $[[7,1,3]]$}
  \end{subfigure}
  \caption{Comparison of separated degenerate decoding of $X$ and $Z$ errors versus joint degenerate decoding (DML decoding).}
  \label{fig:sim}
\end{figure}

Next, we compare the performance of the \emph{separate} degenerate decoding method outlined earlier with the \emph{joint} degenerate decoding, referred to as DML decoding, described in the preceding sections. For simulations, we generated 40,000 samples for each $p$, being the depolarizing noise parameter, corresponding to the $n$-qubit depolarizing channel, where the depolarizing channel acts independently on the $n$ qubits. The performance of both decoding techniques is evaluated by the logical error rate, where a logical error is declared if the estimated $\hat{e}$ differs from the true $e$ in terms of the coset, specifically if $\hat{e} e \notin S$.

{In Figure~\ref{fig:sim}, we present the results of this simulation for the 4-qubit code in subfigure (a) and the 7-qubit code in subfigure (b). \ES{DML decoding outperforms the separate degenerate decoding for both codes.} However, for small values of $p$, the depolarizing noise parameter, the difference is negligible. In the regime of small errors, separate decoding offers a viable alternative with less complexity compared to joint decoding, without significant loss in performance. It is 
worth mentioning that joint NDML decoding for these two codes performs similarly to DML decoding. However, the difference between DML and NDML decoding techniques becomes more apparent for codes with low-weight stabilizer generators, such as quantum LDPC codes \cite{fuentes2021degeneracy}.}

\subsection{Decoding complexity bounds and comparison}

The sum-product {Viterbi} algorithm on the minimal multi-goal-trellis $T=(
\cV,\cE)$ representing the joint-code $\jc[N] = (N,S)$ requires $|\cE|$ multiplications and $|\cE|-|\cV|+1$ additions and has complexity $2|\cE|-|\cV|+1$ operations with real numbers. Now, let's derive upper bounds for $|\cV|$, $|\cE|$, and $2|\cE|-|\cV|$.

The minimal multi-goal-trellis $T$ is biproper, hence for the quaternary code $\jc[N]$ the numbers $|\cV_{t-1}|$ and $|\cV_t|$ can differ by at most a factor of $4$. Since $|\cV_0|=1$ and $|\cV_n|=2^{n+k}$ we obtain the following upper bound,
\begin{equation}\label{bound2}
\log_4 |\cV_t| \le \min \{t,n+k-t\}.
\end{equation} 
To obtain an upper bound on $\cV$, we take an $[[n,k]]$ stabilizer code for which the trellis $T$ satisfies the bound \eqref{bound2} with equality for all $t$. \ES{We call such codes, \emph{maximum complexity (MC) codes}. For example, the $[[4,2]]$ code from Example 1 is an MC code. }

From \eqref{bound2} it follows that an MC $[[n,k]]$ code with even $n+k$ has the maximum $|\cV_t|=2^{n+k}$ for $t = (n+k)/2$, hence, the code reaches the BCJR-Wolf bound \eqref{Wolf_bound} for this $t$. An  $[[n,k]]$  code with odd $n+k$ can not be an MC code since the point of maximum $(n+k)/2$ in \eqref{bound2} is not an integer. 

To get an upper bound for trellis complexity, we take the parameters of an MC code and state the following.
\begin{theorem}[Upper bound on trellis complexity]\label{thm:complexity}
	The minimal multi-goal-trellis $T=(\cV,\cE)$  of an $[[n,k]]$ stabilizer code satisfies the following  upper bounds:
 \begin{align}
 |\cV| &\le \frac{1}{3}\left( 5\cdot 2^{n+k} -2^{2k} -1 \right) < \frac{5}{3}2^{n+k},\label{eq:upV}\\
|\cE| &\le \frac{4}{3}\left(  2^{n+k+1} -2^{2k} -1 \right) < \frac{8}{3}2^{n+k},\label{eq:upE}\\
2|\cE| - |\cV| &\le \frac{1}{3}\left( 11\cdot 2^{n+k} -7\cdot 2^{2k} -7 \right) < \frac{11}{3}2^{n+k}\label{eq:upcomp}.	
 \end{align}
\end{theorem} 
\begin{proof} 
To obtain these upper bounds, we use the exact values of the trellis complexities $|\cV|$ and $|\cE|$ for an MC $[[n, k]]$ codes with even $n+k$. Therefore, they are upper bounds to any other $[[n,k]]$ code. The cardinality of $\cV$ follows from \eqref{bound2} by summing $|\cV_t|$ for all $i=1,\dots,n$.
Since {$|\cV_t|$} always grows for $t=0,\dots,(n+k)/2$, we obtain $|\cE_t| = |\cV_t|$ for all  $t=0,\dots,(n+k)/2$. Moreover, for  $t=(n+k)/2,\dots,n$, {$|\cV_t|$} always decreases, hence $|\cE_{t}| = |\cV_{t-1}|$ for all $t=(n+k)/2+1,...n$. The cardinality $|\cE|$ is computed by summing $|\cE_t|$ for all $t=1,\dots,n$. Finally, the bound on $2|\cE| - |\cV|$ is obtained by using the exact values of $|\cV|$ and $|\cE|$ computed above.
\end{proof}
\ES{Let us compare the impact of using the minimal multi-goal trellis in
the degenerate decoding: without trellis, we have a complexity of  $n\cdot 2^{n+k}$, whereas with trellis, we get $2|\mathcal{E}|-|\mathcal{V}| < \frac{11}{3}\cdot 2^{n+k}$. In summary, utilizing the trellis results in
an improvement of the decoding complexity of over $3n/11$.}
{Notice, that the trellis complexity of a particular code can be much less than the one given by the upper bounds in Theorem~\ref{thm:complexity}. In this case, the gain of using the minimal trellis will be much larger.}  

We now investigate the case of CSS codes. The minimal multi-goal-trellises $T_X$ and $T_Z$ are biproper, hence for the binary joint codes $\jc[N]_X$ and $\jc[N]_Z$, the vertex cardinalities $|\mathcal{V}_{t-1}|$ and $|\mathcal{V}_t|$ can differ by at most a factor of $2$. The number of $X$- and $Z$-stabilizers can vary, resulting in either a larger $X$-stabilizer trellis and a smaller $Z$-stabilizer trellis, or vice versa. {Without loss of generality,} we assume that the number of $X$- and $Z$- stabilizers are the same and therefore equal to $\frac{n+k}{2}$. From now on, we focus on the $T_X$ trellis; however, everything holds for the $T_Z$ trellis as well. Since $|\mathcal{V}_{X,0}|=1$ and $|\mathcal{V}_{X,n}|=2^{k}$, we obtain the following upper bound:
\begin{equation}\label{bound21}
\log_2 |\mathcal{V}_{X,t}| \le \min \{t,n+k-t\}.
\end{equation} 
In order to obtain an upper bound for $|\mathcal{V}_X|$ we take an $[[n,k]]$ CSS code for which the trellis $T_X$ satisfies the bound \eqref{bound21} with equality for all $t$. An example of such code is the $[[4,2,2]]$ stabilizer code from Example~1. With similar arguments as before we derive the following Corollary.

\begin{corollary}[Upper bound on the CSS trellis complexity]
	The minimal multi-goal-trellises $T_X=(\mathcal{V}_X,\mathcal{E}_X)$ and $T_Z=(\mathcal{V}_Z,\mathcal{E}_Z)$ of the 
 $X$-stabilizers and $Z$-stabilizers respectively, of an $[[n,k]]$ CSS stabilizer code,  satisfy the following  upper bounds:
 \begin{align}
 |\mathcal{V}_X| &\le 3\cdot 2^{\frac{n+k}{2}} -2^k - 1 < 3 \cdot 2^{\frac{n+k}{2}}, \label{eq:upVcss}\\
|\mathcal{E}_X| &\le 2(2^{\frac{n+k}{2}+1} - 2^{k} - 1) <2\cdot  2^{\frac{n+k}{2}},\label{eq:upEcss}\\
2|\mathcal{E}_X| - |\mathcal{V}_X| &\le 5\cdot 2^{\frac{n+k}{2}}-3\cdot 2^k - 3 < 5\cdot 2^{\frac{n+k}{2}}  \label{eq:upcompcss}.	
 \end{align}
{Similarly for $|V_Z|$ and $|E_Z|$. Therefore, the total complexity is bounded by,}
\begin{align}
   { (2|\mathcal{E}_X| - |\mathcal{V}_X|) + (2|\mathcal{E}_Z| - |\mathcal{V}_Z|)  < 10\cdot 2^{\frac{n+k}{2}}  }.
\end{align}
\end{corollary} 
\begin{proof} Due to the similarity to the proof of Theorem~\ref{thm:complexity}, this proof is omitted
\end{proof}

\ES{Similarly to our previous analysis, we now compare the complexity of DML in three different scenarios: first, without using a trellis; second, using the ``joint'' minimal multi-goal trellis for joint decoding of the $X$- and $Z$-stabilizers as introduced in Section~\ref{sec:tdegen}; and third, using ``separate'' minimal multi-goal trellises for decoding the $X$- and $Z$-stabilizers separately, as discussed in Section~\ref{sec:dmlcss}:
\begin{itemize}
\item Without trellis: $n \cdot 2^{n+k}$.
\item ``Joint'' minimal multi-goal trellis decoding: $$(2|\mathcal{E}| - |\mathcal{V}|) < \frac{11}{3} \cdot 2^{n+k}.$$
\item ``Separate'' minimal multi-goal trellis decoding: $$(2|\mathcal{E}_X| - |\mathcal{V}_X|) + (2|\mathcal{E}_Z| - |\mathcal{V}_Z|) < 10 \cdot 2^{\frac{n+k}{2}}.$$
\end{itemize}
}

\ES{
Decoding CSS codes separately significantly reduces the decoding complexity compared to decoding without a trellis by more than $\frac{n}{10} \cdot 2^{\frac{n+k}{2}}$. Additionally, compared to joint multi-goal trellis decoding, separate decoding reduces complexity by more than $\frac{11}{30} \cdot 2^{\frac{n+k}{2}}$. While this substantial reduction in complexity may come at the cost of some performance loss, separate decoding remains the most commonly used method in the literature.}

\section{Conclusions}
In conclusion, our investigation unfolds in two parts. We provided, a comprehensive explanation of the trellis-based decoding method for quantum stabilizer codes, we demonstrated that properties described in previous literature can be derived directly from rectangular coding theory. We introduce three approaches for constructing stabilizer code trellises: utilizing stabilizer group generators, employing normalizer generators, and a method for merging twin nodes in the trivial trellis.

In the second part, we demonstrate that the complexity of degenerate decoding for quantum stabilizer codes of length $n$ can be reduced significantly by utilizing the proposed minimal multi-goal trellis. This trellis represents all cosets of the stabilizer group $S$ within the normalizer $N$. We present three distinct approaches for constructing the minimal trellis, each offering unique advantages. To derive these results, we establish several useful properties of the multi-goal trellis. Additionally, we tailor a trellis construction specifically for CSS codes, enabling degenerate decoding using \(X\) and \(Z\) generators independently. Lastly, complexity bounds are established for the proposed DML decoder with the minimal multi-goal trellis.

\ES{In our work we have followed the Hamming approach to coding theory. The memoryless depolarising channel, see Section \ref{sec:depol}, plays a central role in the Hamming approach. In the Shannon approach, which is of course also central to any proof of a capacity theorem, a sequence of capacity-achieving codes is to be constructed for a given quantum channel.  The construction of such capacity-achieving codes for general channels is an important open problem. Even for classical memoryless channels, this task cannot be solved by Turing machines, i.e. there is no Turing machine that receives a memoryless discrete channel as input and then computes a sequence of capacity-achieving codes for this channel \cite{boche2022turing}. Even the calculation of important parameters, such as the capacity-achieving input distributions, is not possible with the help of Turing machines \cite{boche2023algorithmic,lee2023computability}.  Ning Cai has always been very interested in these questions and also in questions concerning practically relevant channel sets for which the corresponding tasks can be solved algorithmically \cite{li2019blahut,li2020computing}. The Hamming approach seems to be very promising for this direction.   }

\balance

\begin{credits}
\subsubsection{\ackname}
The authors acknowledge the financial support by the Federal Ministry of Education and Research of Germany in the programme of “Souverän. Digital. Vernetzt.”. Joint project 6G-life, project identification number: 16KISK002. Holger Boche was supported in part by the Bundesministerium für Bildung und Forschung (BMBF) through the grants 16KISQ093 (QUIET), 16KIS1598K (QuaPhySI), 16KISQ077(QDCamNetz) and 16KISR027K (QTREX).
The authors acknowledge the financial support by the Federal Ministry of Education and Research of Germany (BMBF) in the joint project 6G-life with the project identification number 16KISK002. Evagoras Stylianou was supported in part by  the BMBF project QDCamNetz with the project identification number 16KISQ077. 
\end{credits}
\bibliographystyle{splncs04}
\bibliography{refs}
\end{document}